\documentstyle[preprint,tighten,epsfig,aps,mathbbol,floats,amsmath]{revtex}

\newcommand{\reff}[1]{(\ref{#1})}

\def\gt{\tilde{g}}

\newcommand{\fd}{fluc\-tu\-a\-tion-dis\-si\-pa\-tion }
\def\be{\begin{equation}}
\def\ee{\end{equation}}

\def\bea{\begin{eqnarray}}
\def\eea{\end{eqnarray}}

\def\rmd{{\rm d}}
\def\rme{{\rm e}}
\def\e{\epsilon}
\def\dpar{\partial}
\def\p{\Phi}
\def\t{\tau}
\def\s{\sigma}
\def\ie{i.e.,}
\def\eg{e.g.,}
\def\tm{\tau_m}
\def\L{{\Lambda}}
\def\La{{\mathcal L}}
\def\Li{{\rm Li}}
\newcommand{\eq}{{\rm eq}}
\newcommand{\pp}{{\boldsymbol{\pi}}}
\newcommand{\pt}{{\tilde\Phi}}
\newcommand{\st}{{\tilde\s}}
\newcommand{\ppt}{{\tilde\pp}}
\newcommand{\bt}{{\overline\theta}}
\def\x{{\bf x}}
\def\q{{\bf q}}

\renewcommand{\d}{\delta}
\newcommand{\F}{{\mathcal F}}
\def\k{\kappa}
\def\htit{$h\;$}
\def\O{{\mathcal O}}
\begin{document}

\draft

\title{Slow dynamics in critical ferromagnetic vector models relaxing
from a magnetized initial state}
\author{Pasquale Calabrese${}^{1,2}$ and Andrea Gambassi${}^{3,4}$}
\address{$^1$Institute for Theoretical Physics, Amsterdam University,
  Valckenierstraat 65, 1018 XE Amsterdam, The Netherlands.}
\address{$^{2}$Dipartimento di Fisica dell'Universit\`a di Pisa and INFN, 
Pisa, Italy.}
\address{$^{3}$Max-Planck-Institut f\"ur Metallforschung, Heisenbergstr. 3,
  D-70569 Stuttgart, Germany.}
\address{$^{4}$Institut f\"ur Theoretische und Angewandte Physik,
  Universit\"at Stuttgart, Pfaffenwaldring 57, D-70569 Stuttgart, Germany.}

\date{October 10, 2006}

\maketitle

\begin{abstract} 

Within the universality class
of ferromagnetic vector models with $O(n)$ symmetry and purely 
dissipative dynamics, we study the non-equilibrium critical relaxation 
from a magnetized initial state.
Transverse correlation and response functions are exactly 
computed for Gaussian fluctuations and in the limit of infinite number $n$
of components of the order parameter.
We find that the fluctuation-dissipation ratios (FDRs) for
longitudinal and transverse modes differ already at the Gaussian level. 
In these two exactly solvable cases we completely describe the
crossover from 
the short-time to the long-time behavior, corresponding to a
disordered and a magnetized initial condition, respectively.
The effects of non-Gaussian fluctuations on 
longitudinal and transverse quantities 
are calculated in the first order in 
the $\e$-expansion ($\e=4-d$) and reliable three-dimensional 
estimates of the two FDRs are obtained.

\end{abstract}

\newpage

\section{Introduction}

Consider a ferromagnetic system with vector order parameter
$\p_i(\x,t)$, $i=1,\ldots,n$, a critical point at temperature 
$T_c$ and at vanishing external magnetic field ${\bf h}_c=0$, and
prepare it in some initial configuration (which might correspond to an
equilibrium state at a given temperature $T_0$ and magnetic field 
${\bf h}_0$). 
At time $t=0$ bring the system in contact with a thermal bath of temperature 
$T_b\neq T_0$ in the presence of a magnetic field ${\bf h}_b$. 
The ensuing relaxation process is 
characterized by some {\it equilibration time} $t_{\rm eq}$ 
so that for $t \gg t_{\rm eq}$ equilibrium is
attained, whereas for $0 < t \ll t_{\rm eq}$, 
the evolution depends on the specific initial condition.
Upon approaching the critical point
$T_b = T_c$, ${\bf h}_b = {\bf h}_c = 0$ 
the equilibration time diverges and therefore
equilibrium is never achieved. 
One-time quantities (like mean-energy, magnetization, etc.) typically
display a slow relaxation (power-law) towards their equilibrium values, even in the 
extreme situation in which $t_{\rm eq}$ is infinite. 
Consequently, the natural objects to monitor the non-equilibrium
evolution of the system are two-time
quantities, like the time-dependent correlation function  of some
local observable 
$\O$ which is 
given by $C_{\bf x}^\O(t,s)= \langle \O({\bf x},t)\O({\bf 0},s)\rangle$, 
($\langle \ldots \rangle$ stands for the average over the stochastic dynamics) 
and the linear response $R_{\bf x}^\O(t,s)$ to an external 
field $h_\O$ conjugate to $\O$.  
$R_{\bf x}^\O(t,s)$ is defined by
$R_{\bf x}^\O (t,s)=\delta\langle \O({\bf x},t)\rangle/ \delta h_\O (s)|_{h_\O=0}$,
where $h_\O$ is applied at time $s > 0$ at the point ${\bf x}=0$. 
For $t > s \gg t_\eq$ correlation and response functions take their 
equilibrium forms  --- depending only on $t-s$ --- and they are related by the {\it \fd theorem} (FDT)
\be
T_b R^\O_{\bf x}(t\gg t_\eq, s\gg t_\eq) =  
\dpar_s C^\O_{\bf x}(t\gg t_\eq,s\gg t_\eq)\;.
\label{eqFDT}
\ee
This is no longer true in the so-called aging regime: correlation and response 
functions have a non-trivial (homogeneous) 
dependence on both $t$ and $s$, even for 
very large times. 
In particular their decays as functions of $t$ become slower upon
increasing $s$, the age of the system. 
This is why this evolution is referred to  
as aging.

The FDT suggests the introduction of the so-called {\it \fd ratio} 
(FDR)~\cite{ck-93,ckp-94}:
\be
X_{\bf x}^\O(t,s)=\frac{T_b\, R_{\bf x}^\O(t,s)}{\dpar_s C_{\bf x}^\O(t,s)}
\label{dx}
\ee
which can be used in order to understand the exten to which a 
system is evolving
out of equilibrium.
The asymptotic value of the FDR 
\be
X^\infty_\O=\lim_{s\to\infty}\lim_{t\to\infty}X_{{\bf x}=0}^\O(t,s)
\label{xinfdef}
\ee
is a useful quantity in the description
of systems with slow dynamics, since  $X^\infty_\O=1$ whenever 
the aging evolution is interrupted and  
the system crosses over (for $t>s\gg t_{\rm eq}$) to equilibrium dynamics. 
Conversely,
$X^\infty_\O\neq 1$ is a signal of an asymptotic non-equilibrium dynamics.

Within the field-theoretical approach to critical dynamics
it is more convenient to focus on the
behavior of observables in {\it momentum} space. Accordingly,
hereafter we mainly consider the momentum-dependent response $R_{\bf q}^\O(t,s)$
and correlation $C_{\bf q}^\O(t,s)$ functions, 
defined as the Fourier transforms of $R_\x^\O (t,s)$ and $C_\x^\O(t,s)$, 
respectively.
In momentum space it is natural to introduce a 
quantity with the same role as $X_{\bf x}^\O(t,s)$~\cite{cg-02a1}:
\be
{\cal X}_{\bf q}^\O(t,s)=\frac{T_b R_{\bf q}^\O(t,s)}{\dpar_s C_{\bf q}^\O(t,s)} \;.
\label{Xq} 
\ee
It has been argued that the long-time limit 
\be
{\cal X}^\infty_\O\equiv 
\lim_{s\rightarrow\infty}\lim_{t\rightarrow\infty} {\cal X}_{{\bf q}=0}^\O(t,s)
\label{eq} \;
\ee
has exactly the same value as $X^\infty_\O$~\cite{cg-02a1} (see also Ref.~\cite{sp-05}).

Aging was known to occur in disordered and complex systems (see, e.g., Ref.~\cite{review}) and only recently attention has been focused on 
simpler critical systems such as ferromagnets, whose universal
behaviors can be rather
easily studied by using powerful theoretical methods and which might provide
insight into more general cases~\cite{cg-rev} (see Ref.~\cite{g-05}
for a pedagogical introduction).  
In particular it was argued that $X^\infty$ is a
{\it universal amplitude ratio} for quenches to $T_c$ \cite{gl-00c}. 
In this respect several dynamic universality classes were investigated
and the scaling forms the of response and correlation functions,
together with the associated FDRs 
were calculated by means of exact, approximate, or numerical 
methods~\cite{gl-00i,cg-02a2,cg-02rim,ph-02,lsi,cg-03,mbgs-03,sdc-03,ch-03,ak,gkf-03,cg-04,hs-04,clz-04,pg-04,ft-05,gspr-05,as-05,cl-05,p-04,bp-06,cgk-06}
(see Ref.~\cite{cg-rev} for a review).

In this paper we consider the purely dissipative dynamics
(Model A, according to the classification of Ref.~\cite{HH}) of the $O(n)$ 
symmetric model relaxing from a magnetized state (in a sense that will be 
specified below).
Previous works \cite{bj-76,bej-79,cgk-06} mainly focused on 
the case $n=1$ --- corresponding to the Ising universality class --- because
the dynamics of actual systems with vector order parameter and
$O(n>1)$ symmetry typically possesses conservation laws which are {\it
not} accounted for by Model A.
For example, an isotropic ferromagnet 
(antiferromagnet) is correctly described by the so-called Model J (G) 
universality class~\cite{HH}.
Nonetheless, 
we consider this {\it unphysical} model because it is probably the simplest 
example in which the dynamics is affected by two different fluctuation 
modes. In fact, consider the 
relaxation from an initial state with a non-vanishing magnetization 
${\bf M}_0$. In the absence of external fields the magnetization will
evolve as $M(t)$ in the direction ${\bf \widehat M}_0 \equiv {\bf M}_0/M_0$.
The time evolution of the
component of the order parameter parallel to 
${\bf M}(t)$ is clearly different from the evolution of the $n-1$
components perpendicular to it, which are characterized by  
a residual $O(n-1)$ symmetry.  
In what follows the mode $\s(\x,t) = \p(\x,t)\cdot{\bf \widehat M}_0 - M(t)$ 
parallel to ${\bf\widehat M}_0$
is referred to as {\it longitudinal},
whereas the $n-1$ perpendicular modes $\pp(\x,t) = \p(\x,t) - {\bf
\widehat M}_0 [{\bf \widehat M}_0\cdot\p(\x,t)]$ 
are referred to as {\it transverse}. 
They are characterized by different responses and correlations, with  FDRs:
\be
X^\infty_\s=\lim_{s\to\infty}\lim_{t\to\infty} 
\frac{T_b R_{\q=0}^\s(t,s)}{\dpar_s C_{\q=0}^\s(t,s)}\,,\qquad
X^\infty_\pp=\lim_{s\to\infty}\lim_{t\to\infty} 
\frac{T_b R_{\q=0}^\pp(t,s)}{\dpar_s C_{\q=0}^\pp(t,s)}\,.
\ee
In spite of its unphysical nature, the purely dissipative dynamics of the
$O(n)$ model has
a rich behavior which is easily
accessible to numerical simulations and possibly captures some of the
features of more realistic dynamics. 
The model has already been studied by Fedorenko and Trimper
\cite{ft-05} in a dimensional expansion close to the 
lower critical dimension --- an approach 
that is complementary to the one adopted here.

The paper is organized as follows. In Sec.~\ref{sec2} we introduce the
model and we discuss the general scaling forms for the 
correlation and response functions of longitudinal and transverse modes 
after a quench from a magnetized state. 
In Sec.~\ref{sec-Gaux} we solve the model within the Gaussian approximation. 
In Sec.~\ref{Secninf} we derive the exact transverse response and 
correlation functions in the limit of infinite number of components of the 
order parameter.  
In these two exactly solvable cases we describe completely the
crossover from 
the short-time to the long-time behavior, corresponding to a
disordered and a magnetized initial condition, respectively.
In Sec.~\ref{sec-onel} we account for non-Gaussian fluctuations 
up to first-order in the $\e$-expansion ($\e=4-d$) and from the
resulting expressions of the  response and correlation functions 
we calculate the associated FDR.
Finally in Sec.~\ref{sec-con} we summarize and discuss our results.

\section{The model}
\label{sec2}

The purely dissipative dynamics of a $n$-component field 
$\Phi=(\Phi_1,\dots,\Phi_n)$ 
can be specified in terms of the stochastic Langevin equation
\be
\label{lang}
\dpar_t \p_i ({\x},t)=-\Omega
\frac{\delta {\cal{H}}[\p]}{\delta \p_i(\x,t)}+\xi_i(\x,t) \; ,
\ee
where $\Omega$ is the kinetic coefficient,
$\xi_i(\x,t)$ a set of zero-mean stochastic Gaussian noises with
\be
\langle \xi_i(\x,t) \xi_j(\x',t')\rangle= 2 \delta_{ij} \Omega \, \delta(\x-\x') \delta (t-t'),
\ee
and ${\cal H}[\p]$ is the static Hamiltonian.
Near the critical point, ${\cal{H}}[\p]$ may be assumed of the $O(n)$-symmetric
Landau-Ginzburg form
\be
{\cal H}[\p] = \int \rmd^d x \left[
\frac{1}{2} (\nabla \p )^2 + \frac{r_0}{2} \p^2
+\frac{g_0}{4!} (\p^2)^2 \right] ,\label{lgw}
\ee
where $r_0$ is a parameter that has to be tuned to a critical value
$r_{0,c}$ in order to
approach the critical temperature $T=T_c$ ($r_{0,c}=0$,
within the analytical approach discussed below), and $g_0>0$ is the bare
coupling constant of the theory.

Correlation and response functions of a field which evolves according
to  the Langevin equation~\reff{lang} can be obtained by means of the
field-theoretical action~\cite{zj,bjw-76}
\be
S[\p,\tilde{\p}]= \int_0^\infty \rmd t \int \rmd^dx
\left[\tilde{\p} \dpar_t\p +
\Omega \tilde{\p} \frac{\delta {\mathcal{H}}[\p]}{\delta \p}-
\tilde{\p} \Omega \tilde{\p}\right]\,,\label{mrsh}
\ee
where $\pt(\x,t)$ is an auxiliary field, conjugate to
the external field ${\bf h}$ in such a way that
${\cal H}[\p,{\bf h}] = {\cal H}[\p] - \int \rmd^d x {\bf h}\cdot\p$.

The effect of a macroscopic initial condition
$\p_0(\x)\equiv\p(\x,t=0)$ may be accounted for by
averaging over the initial configuration
with a weight $\rme^{-H_0[\p_0]}$, where~\cite{jss-89,jan-92}
\be
H_0[\p_0]=\int\! \rmd^d x\, \frac{\tau_0}{2}[\p_0(\x)-{\bf M}_0]^2
\label{idist}
\ee
specifies an initial state with Gaussian distribution 
of the order parameter fluctuations, short-range
correlations proportional to
$\tau_0^{-1}$, and spatially constant average order
parameter ${\bf M}_0 = \langle \p_0(\x)\rangle$.
In the experimental protocols we have in mind the initial state with
${\bf M}_0\neq 0$ can be obtained by equilibrating the system
for $t<0$ in the presence of an external
field ${\bf h}_0$ and at generic temperature $T_0$.  
The resulting state can be described by Eq.~\reff{lgw} with external
field ${\bf h}_0$, leading, far enough from the critical point, to a
Gaussian distribution of the order parameter 
such as Eq.~\reff{idist} with  $\tau_0 \sim
g_0 M_0^2$ where $M_0 = (6 |{\bf h}_0|/g_0)^{1/3}$.
However, within the renormalization-group (RG) approach to the problem it has been
shown~\cite{jss-89} that $\tau_0^{-1}$ is an irrelevant variable in
the sense that $\tau_0^{-1}$ affects only the correction to the
leading long-time scaling behavior we are interested in. 
In view of that we fix it to the value $\tau_0^{-1}=0$ from the very beginning
of the calculation. 
It is important to note that the initial state described by 
by the Hamiltonian Eq.~\reff{idist} {\it does not} correspond to a
low-temperature one in zero magnetic field, unless $n=1$~\cite{cgk-06}.
In fact, for $n>1$ the transverse modes are 
critical for all $T<T_c$ (Goldstone theorem, see, e.g.,
\cite{zj}) and the proper $H_0$ takes the form of Eq.~\reff{idist}
only for the longitudinal mode, whereas the most relevant term (in the RG 
sense) for the transverse ones would be 
$\int\! \rmd^d x\, (\nabla \pp_0)^2$ (where $\pp_0(\x,t) \equiv \pp(\x,t=0)$).

In order to account for a non-vanishing mean value of the order
parameter $\langle \p(\x,t)\rangle = {\bf M}(t)$ (we assume that ${\bf M}(t)$
stays homogeneous in space after the quench)
during the time evolution,
it is convenient~\cite{bej-79} to write the
action~\reff{mrsh} in terms of the transverse and longitudinal
fluctuations ($\pp$ and $\s$, respectively) around ${\bf M}(t)$, i.e.
\be
\p(\x,t)=(\s(\x,t)+M(t),\pp(\x,t))\,, \qquad
\pt(\x,t)=(\st(\x,t),\ppt(\x,t))\,.
\ee
With this parameterization $\langle \s(\x,t) \rangle = 0$.
The problem we shall consider below is
therefore the dynamics of fluctuations of the fields
$(\s(\x,t),\pp(\x,t))$ in the time-dependent ``background''
provided by $M(t)$.
For later convenience we also introduce the scaled magnetization
$m(t)$:
\be
m^2 \equiv g_0\frac{M^2}{2}\,,
\label{mresc}
\ee
so that a perturbative expansion in $g_0$ leads to a finite value for
$m_0 \equiv m(t=0)$.
The resulting action in terms of $\s, \st,\pp,\ppt$ may be written as
\be
S=\int_{0}^\infty \rmd t \int \rmd^dx  ({\cal L}_0+{\cal L}_1+{\cal L}_2)\,,
\label{S123}
\ee
where
\bea
{\cal L}_0&=&-\st \Omega \st +\st
\left(\dpar_t+\Omega \left[-\nabla^2+r_0+m^2(t)\right]\s\right)\nonumber\\
&&-\ppt \Omega \ppt +\ppt
\left(\dpar_t+\Omega \left[-\nabla^2+r_0+\frac{m^2(t)}3\right]\right)\pp
\,,\label{L0}\\
{\cal L}_1&=&\Omega\left[\sqrt2 m(t) \left(
\frac{\sqrt{g_0}}{2} \st \s^2+ \frac{\sqrt{g_0}}{6} \st\pp^2
+\frac{\sqrt{g_0}}{3} \s \ppt\pp\right)\right.\nonumber\\
&&+\left.
\frac{g_0}{6} \st \s^3+\frac{g_0}6 \st\s\pp^2+\frac{g_0}6 \ppt \pp \pp^2
+\frac{g_0}6 \s^2\ppt\pp
\right]\,,\label{L1}\\
{\cal L}_2&=&\st\left(\dpar_t+
\Omega\left[r_0+\frac{1}{3}m^2(t)\right]\right)
\sqrt{\frac{2}{g_0}}m(t) \equiv \st h_{\rm eff}(t)\label{L2}\,.
\eea
We split up the action $S$ so that ${\cal L}_0$ is the Gaussian part,
${\cal L}_1$ contains the interaction vertices and ${\cal L}_2$ gives
the coupling to the effective magnetic field $h_{\rm eff}(t)$, due to a nonzero
$M(t)\propto m(t)$,
acting on $\s(\q=0,t)$.
The presence of $m(t)\neq 0$ breaks 
the original $O(n)$ symmetry of $S$ in the fields $\p$ and $\pt$ (see
Eq.~\reff{mrsh}), leaving only a reduced $O(n-1)$ symmetry in the
transverse field $\pp$ and $\ppt$ (to all orders in perturbation theory).

\begin{figure}[tb]
\centerline{\epsfig{width=11truecm,file=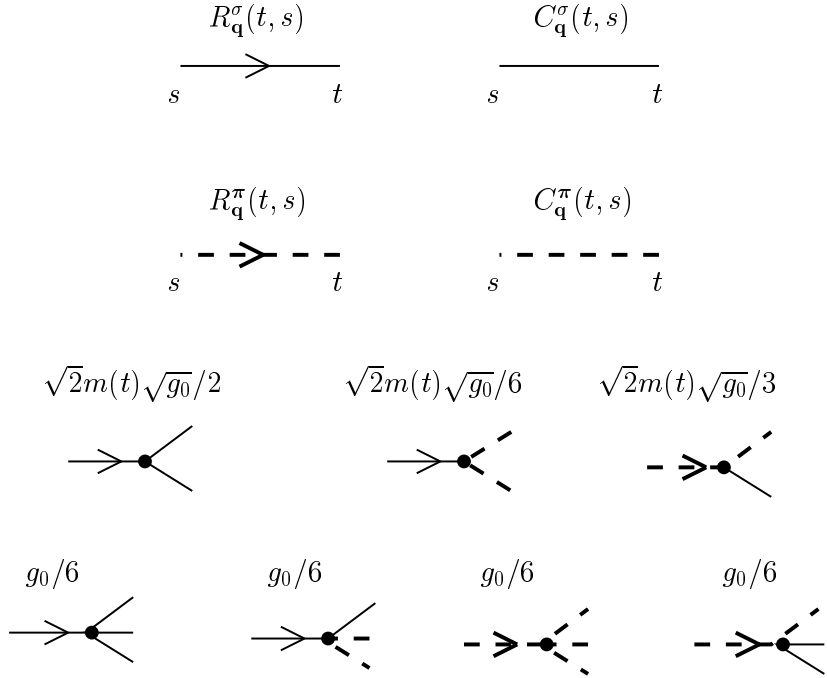}}
\caption{Elements of the diagrammatic representation of the
perturbation theory.}
\label{Frules}
\end{figure}

Following standard methods \cite{zj,bjw-76} the response and
correlation functions may be obtained by a perturbative expansion of
the functional weight $\rme^{-(S[\p,\pt]+H_0[\p_0])}$ in terms of the
coupling constant $g_0$. The propagators (Gaussian
two-point functions of the fields $\s$ and $\pp$ in momentum space)
are~\cite{jss-89,jan-92}
\bea
\langle \s(\q,t) \st(\q',t')\rangle_0&=& (2\pi)^d \d(\q+\q') R^\s_\q(t,t')\,,\\
\langle \s(\q,t) \s(\q',t') \rangle_0&=& (2\pi)^d \d(\q+\q') C^\s_\q(t,t')\,,\\
\langle \pp_i(\q,t) \ppt_j(\q',t')\rangle_0&=& (2\pi)^d \d_{ij}\d(\q+\q') R^\pp_\q(t,t')\,,\\
\langle \pp_i(\q,t) \pp_j(\q',t') \rangle_0&=& (2\pi)^d
\d_{ij}\d(\q+\q') C^\pp_\q(t,t')\,,
\eea
where ($q=|\q| $)
\bea
R^\s_\q(t,t')&=& \theta(t-t') \exp\left\{ -\Omega\left[(q^2+r_0)(t-t')+
\int_{t'}^t \rmd t'' m^2(t'')\right]\right\}\,,\label{RLgaux}\\
R^\pp_\q(t,t')&=& \theta(t-t') \exp\left\{ -\Omega\left[(q^2+r_0)(t-t')+
\int_{t'}^t \rmd t'' m^2(t'')/3\right]\right\}\,,\label{RTgaux}\\
C^{\s,\pp}_\q(t,t')&=&2\Omega \int_0^\infty \rmd t'' R^{\s,\pp}_\q(t,t'')
R^{\s,\pp}_\q(t',t'').\label{Cgaux}
\eea
The perturbative expansion can be as usual organized in terms of
Feynman diagrams. 
The response propagators are given by Eq.~\reff{RLgaux} for the
longitudinal mode and by Eq.~\reff{RTgaux} for the transverse one, and
they are represented as directed solid and dashed lines, respectively,
where the arrow points towards $t$, the larger of the two times. The
correlation propagators are given, instead, by Eq.~\reff{Cgaux} and
are represented as solid and dashed lines for longitudinal and
transverse modes, respectively.
The interaction vertices are contained in ${\cal
L}_1$: Four of them are standard
time-independent quartic vertices $\propto \Omega g_0$ whereas the
remaining three are  time-dependent cubic ones $\propto
\Omega\sqrt{g_0} m(t)$, due to a non-zero magnetization
$m(t)$. Propagators and vertices are shown in Fig.~\ref{Frules}.
The contribution of $h_{\rm eff}(t)$ cancels for a suitable choice
of $m(t)$. The evolution equation for the magnetization $m(t)$ can
be obtained by solving the equation of motion  $\langle \delta
S/\delta \st(\x,t)\rangle|_{\st = 0} = 0$:
\be
\left[\dpar_t+ \Omega\left(r_0+\frac{1}{3}m^2(t)\right)\right] m(t)  +
\Omega\frac{g_0}{2} m(t)\langle\s^2+ \pp^2/3\rangle
+ \frac{\Omega g_0^{3/2}}{6\sqrt2}\langle \s^3+\s\pp^2\rangle = 0\;.
\label{eqmg}
\ee
To the lowest order [hereafter $(\rmd q)=\rmd^d q/(2\pi)^d$ and $\Omega=1$]
\bea
\langle \s^2(\x,t) \rangle &=& \int (\rmd q) C^\s_\q(t,t) + O(g_0)\;,
\quad\quad
\langle \s^3(\x,t)\rangle = O(\sqrt{g_0})\,,\\
\langle \pp^2(\x,t) \rangle &=& (n-1)\int (\rmd q) C^\pp_\q(t,t) + O(g_0)\;,
\quad\quad
\langle \s\pp^2(\x,t)\rangle = O(\sqrt{g_0})\,,
\eea
and therefore up to one loop and at the critical point $r_0 = 0$,
Eq.~\reff{eqmg} becomes
\be
0 = \dpar_t m(t) +\frac{m^3(t)}{3}  + \frac{g_0}{2}\int
(\rmd q) \left[ C^\s_\q(t,t)+\frac{n-1}3 C^\pp_\q(t,t)\right] + O(g_0^2) \,.
\label{eqmo}
\ee

\subsection{Scaling forms}
\label{sec-scaling-forms}
When a critical system is quenched from the high temperature phase
to the critical point response and correlation functions display the
scaling forms~\cite{jss-89,cg-rev}
\bea
R_{\q=0}(t,s) &=& A_R\, (t-s)^a(t/s)^\theta F_R(s/t)\; ,
\label{scalRm0}\\
C_{\q=0}(t,s) &=& A_C\,s(t-s)^a(t/s)^\theta F_C (s/t)\; ,
\label{scalCm0}
\eea
where $a = (2-\eta-z)/z$, $z$ is the dynamical critical exponent,
$\eta$ the anomalous dimension of the field, and $\theta$ is a
genuine non-equilibrium exponent~\cite{jss-89}. $A_R$ and $A_C$ are
non-universal amplitudes which are fixed by the condition
$F_{R,C}(0)=1$. With this normalization $F_{R,C}$ are universal and
$X^\infty$ turns out to be a 
{\it universal amplitude ratio} 
$X^\infty = A_R/[A_C(1-\theta)]$~\cite{gl-00c,cg-rev}.
Although we shall focus mainly on the
case $\q=0$, the generalization of these scaling forms to
non-vanishing $\q$ amounts to the introduction of an additional
scaling variable $y = A_\Omega \Omega q^z (t-s)$. $A_\Omega$ is a
dimensional non-universal constant which can be fixed according to
some specified condition. This scaling behavior gets
modified when the critical system evolves from
a state with $M_0\neq0$. The magnetization scales according to~\cite{jss-89}
\be
m(t) = A_m m_0 t^{a+\theta} \F_M(B_m m_0 t^\k)
\label{scalm0}
\ee
where $\k = \theta + \beta\delta/(\nu z)-1$ and standard notation
for critical exponents has been used. The non-universal amplitudes
$A_m$ and $B_m$ can be determined, \eg\ by imposing $\F_M(0)= 1$ and
$\F''_M(0)= -1$ \cite{cgk-06}.
From Eq.~\reff{scalm0} one sees that the effect of a non-vanishing
initial magnetization $m_0$ is the introduction of an additional
macroscopic time scale $\tau_m$ into the problem, \ie\ $\tau_m =
(B_m m_0)^{-1/\k}$ and of an additional associated scaling variable
$u\equiv t/\tau_m$ in the scaling forms. This scaling variable,
naturally appears 
in the scaling forms for the response
[Eq.~\reff{scalRm0}] and {\it connected} correlation functions
[Eq.~\reff{scalCm0}]: 
\bea
R_{\q=0}^i(t,s) &=& A_R\, (t-s)^a(t/s)^\theta {\cal
F}_R^i(s/t,B_m m_0 t^\k)\; ,
\label{scalR}\\
C_{\q=0}^i(t,s) &=& A_C\,s(t-s)^a(t/s)^\theta {\cal F}_C^i
(s/t, B_m m_0 t^\k)\;,
\label{scalC}
\eea
where no new non-universal amplitudes have been introduced. The
index $i=\s,\pp$ indicate the longitudinal and transverse fluctuation
modes, respectively. The resulting functions ${\cal F}_R$ and ${\cal F}_C$ are
universal. For $m_0=0$ the original $O(n)$ symmetry of the model is
restored and therefore transverse and longitudinal response (and correlations)
become identical.

According to Eqs.~(\ref{scalR}) and (\ref{scalC}) a non-vanishing
mean value of the initial magnetization $m_0\neq 0$ affects the
scaling properties of the response and correlation function as soon
as $B_m m_0 t^\k \sim 1$ (\ie\ $t\sim \tau_m$) and in particular
this happens in the long-time limit we are interested in,
characterized by $t\gg s\gg\tau_m$.
This formally corresponds to the case $m_0\rightarrow \infty$, as
opposed to the case previously considered, $m_0=0$. In this limit
one expects the scaling forms~\reff{scalR} and~\reff{scalC} to turn
into:
\bea
R_{\q=0}^\s(t,s) &=& A_R^\s\, (t-s)^a(t/s)^{\theta_\s} F_R^\s
(s/t)\; ,
\label{scalRL}\\
C_{\q=0}^\s(t,s) &=& A_C^\s\,s(t-s)^a(t/s)^{\theta'_\s} F_C^\s
(s/t)\; ,
\label{scalCL}\\
R_{\q=0}^\pp(t,s) &=& A_R^\pp\, (t-s)^a(t/s)^{\theta_\pp}
F_R^\pp (s/t)\; ,
\label{scalRT}\\
C_{\q=0}^\pp(t,s) &=& A_C^\pp\,s(t-s)^a(t/s)^{\theta'_\pp}
F_C^\pp (s/t)\; .
\label{scalCT}
\eea
The non-universal constants $A^i_{R,C}$ are fixed by
requiring $F^i_{R,C}(0) =1$, where $F^i_{R,C}(x)$ are universal
functions related to the large-$v$ behavior of $\F^i_{R,C}(x,v)$.
The exponents $\theta_i$ and $\bt'_i$ are different from $\theta$.
In particular it is not obvious, a priori, whether  $\theta_i$ and
$\theta'_i$ should be expected to be the same, if they are novel
exponents --- as $\theta$ --- or just combinations of known ones.

In Ref.~\cite{cgk-06} we showed, for the Ising model (\ie\
the $O(n)$ model with $n=1$), that
\be
\theta_\s = \theta'_\s = - \frac{\beta\delta}{\nu z}\, .
\label{conclL}
\ee
The RG proof of Ref.~\cite{cgk-06} made no use of the scalar nature
of the order parameter and holds for any purely dissipative model
(\ie\ with Model A dynamics).

For the transverse modes, it has been argued in Ref.~\cite{ft-05}
that
\be
\theta_\pp = \theta'_\pp = - \frac{\beta}{\nu z}\, .
\label{conclT}
\ee

A fundamental difference between the quench from a disordered and a
magnetized state emerges: In the latter case 
the aging properties and the non-equilibrium
behavior of both longitudinal and transverse modes 
are completely described in terms of equilibrium exponents.

As a consequence of the scaling forms Eqs.~\reff{scalRL},
\reff{scalCL}, \reff{scalRT}, and \reff{scalCT} the two limiting FDR
are
\be
X^\infty_\s(m_0\neq0) =\frac{A_R^\s}{A_C^\s(1-\theta_\s)}\,,\qquad
X^\infty_\pp(m_0\neq0)=\frac{A_R^\pp}{A_C^\pp(1-\theta_\pp)}\,,\\
\label{Xinfamprat}
\ee
which turn out to be {\it universal amplitude ratios}~\cite{cgk-06}.

\section{Gaussian Approximation}
\label{sec-Gaux}

The starting point for the analysis of the dynamics resulting from the
action $S$ [see
Eqs.~\reff{mrsh} and~\reff{S123}] 
is the Gaussian approximation in which one 
neglects all the terms which are of degree higher than 2 in the fields
$\pp$ and $\s$, contained in ${\cal L}_1$.
The result of such an approximation is 
exact if the space dimensionality is greater than the upper 
critical dimension $d_u=4$~\cite{zj,cg-rev}.

Within this approximation, the equation of
motion~\reff{eqmo} at the critical point $r_0=0$ becomes
\be
\dpar_t m_G(t) =-\frac{1}{3} m_G^3(t)\,,
\ee
($G$ stands for Gaussian) whose solution is
\be
\label{mgauss}
m_G^2=\frac{1}{\frac{2}{3} t +\frac{1}{m^2_0}}  = (2 t/3)^{-1}
\left(1 + \frac{3}{2 m_0^2 t}\right)^{-1}\,.
\ee
This expression agrees with the scaling behavior~\reff{scalm0}, 
given that $\theta=a = 0$ and 
$\k = \frac{1}{2}$ within the Gaussian model. The non-universal amplitudes
appearing in Eq.~(\ref{scalm0}) are $A_m=1$ and $B_m=\sqrt{2/3}$,
leading to $\tau_m = (B_m m_0)^{-1/\k} = 3/(2m_0^2)$ and
$m_G(t) = m_0(1+t/\tm)^{-1/2}$. For $t \gg \tau_m$, $m_G(t) \sim (2
t/3)^{-1/2}$ and a nonzero $m_0^{-1}$ affects only the
corrections to this leading long-time behavior. In this sense
$m_0^{-1}$ is irrelevant for long times.

Gaussian response and correlation functions (see Eqs.~\reff{RLgaux}
and~\reff{Cgaux}) do not depend on $n$ and therfore the longitudinal
ones are the same as those of the Ising model studied in
Ref.~\cite{cgk-06}, which the reader is referred to for details.
We recall here that the Gaussian FDR turns out to
be $X^\infty_\s=4/5$ \cite{cgk-06}. For the transverse critical
response and correlation functions,
Eqs.~\reff{RTgaux} and~\reff{Cgaux} give ($t>s$)
\bea
R^{\pp,0}_\q(t,s) &=& \rme^{-\q^2(t-s)-\int_s^t \rmd t'
m_G^2(t')/3}=
\left(\frac{s+\tm}{t+\tm}\right)^{1/2}\rme^{-\q^2(t-s)}\,,
\label{Rgauxeq}\\
C^{\pp,0}_\q(t,s)&=&2 \int_0^s \rmd t'
R^{\pp,0}_\q(t,t')R^{\pp,0}_\q(s,t')= 2
\frac{\rme^{-\q^2(t+s)}}{[(t+\tm)(s+\tm)]^{1/2}} \int_0^{s} \rmd t'
(t'+\tm) \rme^{2\q^2 t'}\,.
\label{Cgauxeq}
\eea
In particular, for $\q=0$ one finds
\bea
R^{\pp,0}_{\q=0}(t,s) &=& \left( \frac{s+\tm}{t+\tm}\right)^{1/2}\,,
\label{Rgauxeq0}\\
C^{\pp,0}_{\q=0}(t,s) &=&
\frac{(s+\tm)^2-\tm^2}{[(s+\tm)(t+\tm)]^{1/2}}\,.\label{Cgauxeq0}
\eea
Comparing these results (for $\tm = 0$), with the scaling
forms~\reff{scalRT} and~\reff{scalCT} we can identify $z=2$, $a=0$,
$\theta_\pp=-1/2$ in agreement with standard mean-field exponents 
($\nu=\beta=1/2$ and $\eta=0$) and determine $A_R^\pp = 1$, $A_C^\pp
= 1$, $F_R^\pp (x)=1$, and $F_C^\pp(x)=1$. Accordingly, the asymptotic
FDR is given by (see Eq.~\reff{Xinfamprat})
\be
X^\infty_\pp=\frac{A_R^\pp}{A_C^\pp(1-\theta_\pp)}=\frac23\,,
\ee
which differs from the corresponding quantity for the longitudinal mode
$X^\infty_\s=4/5$.
To our knowledge this is the first case in which the asymptotic FDR
of two local quantities differ already within the Gaussian
approximation. 
In fact in Ref.~\cite{cg-04} we showed that, for a quench from the
disordered phase to the critical point, the Gaussian $X^\infty_\O$ is
always $1/2$, independently of the specific quantity $\O$ it refers
to.
As a consequence an effective temperature $T_{\rm eff} \equiv
T_c/X^\infty$ is not well-defined even within the Gaussian
approximation for the quench from a magnetized
state~\cite{ckp-97,ccy-06,cg-04}.

\begin{figure}[tb]
\centerline{\epsfig{width=\textwidth,file=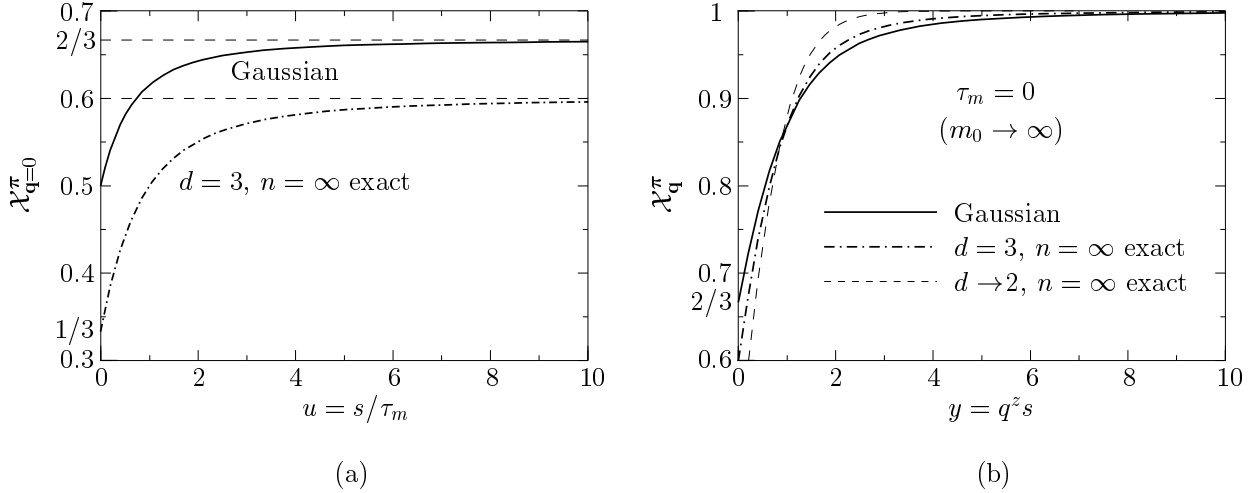}}
\caption{Transverse FDR ${\mathcal X}^\pp_\q$ within the Gaussian
approximation (solid line) and in the exactly solvable limit
$n\rightarrow\infty$ for $d=3$ (dash-dotted line) and $d \to 2$
(dotted line in the right panel).
Left: ${\cal X}^\pp_{\q=0}$ as a function of $s/\tau_m$. 
Right: ${\cal X}^\pp_\q$ for $\tau_m=0$ as a function of $q^z s$.}
\label{Xcombi}
\end{figure}

Let us discuss in more detail the effect of a non-zero magnetization on the
FDR
\be
{\cal X}^\pp_{\q}(t,s) \equiv \frac{R^{\pp,0}_\q(t,s)}{\partial_s
C^{\pp,0}_\q(t,s)} \;.
\ee
From Eqs.~\reff{Rgauxeq0} and~\reff{Cgauxeq0} one finds, as a
function of $u=s/\tm$,
\be
{\cal X}^\pp_{\q=0}(u) = \frac23\left[1+\frac13(1+u)^{-2}\right]^{-1}\,,
\label{Xcrossu}
\ee
which is actually independent of $t$ (a typical property of the Gaussian
approximation) and increases monotonically from the value $1/2$ for
$u=0$, to the value $2/3$ for $u\to\infty$. 
Accordingly, whenever $m_0$ is non zero,  the
transverse FDR approaches $2/3$ 
in the long-time limit (and its
value is indeed independent of the actual $m_0\neq 0$) whereas the
FDR is equal to $1/2$ for $m_0=0$, \ie\ for quenches from a
disordered initial state. The plot of ${\cal X}^\pp_{\q=0}(s,t)$ as a
function of $u = s/\tau_m$ is reported in Fig.~\ref{Xcombi}(a).
%
%
%
%
In the case $\tau_m = \infty$ (corresponding to $m_0=0$) it was found
that the fluctuation mode responsible for aging is the homogeneous one
${\bf q} = 0$, given that for ${\bf q} \neq 0$ and long times ${\cal
X}^\s_{\bf q} = {\cal
X}^\pp_{\bf q} = 1$~\cite{cg-02a1,cg-rev}. The same conclusion was
drawn for the (longitudinal) fluctuations of the Ising model with
$\tau_m=0$ (corresponding to $m_0=\infty$)~\cite{cgk-06}. 
Now for the transverse
modes one finds, from Eqs.~\reff{Rgauxeq} and~\reff{Cgauxeq}, 
as a function of $y=q^2 s$,
\be
{\cal X}^\pp_\q(y)=\frac{4 y^2}{1+4 y^2-\rme^{-2 y}(1+2 y)}\,,
\label{XyGaux}
\ee
which monotonically increases from the value $2/3$ at $y=0$
to $1$ for $y\rightarrow\infty$ [see Fig.~\ref{Xcombi}(b)],
as observed in all the other cases studied so far.

\section{Dynamics of the transverse modes in the limit $n\to\infty$}
\label{Secninf}

Critical transverse response and correlation functions can be analytically 
calculated in the limit of large number of components of the 
order parameter. In fact, assuming that $g_0 \equiv g/n$ (with $g$
independent of $n$), the action~\reff{S123} for large $n$
may be written as ($\Omega=1$)
\be
S(\pp,\ppt;m)=\int \rmd^d x \rmd t\left\{-\ppt  \ppt +\ppt
\left[\dpar_t+\left(-\nabla^2+r_0+\frac{m^2(t)}3\right)\right]\pp+
\frac{g}6 \frac{1}{n}\ppt \pp \pp^2\right\}+O(n^0)\,,
\ee
where $m(t)$ satisfies the equation of motion
\be
\left(\dpar_t+r_0+\frac{m^2(t)}3+\frac{g}6 \frac{\langle\pp^2
\rangle}{n} +O(n^{-1})\right)
m(t)=0\,.
\label{eqmotionninf}
\ee
It is evident that only transverse modes contribute to the leading term of the 
action. On the one hand, this makes the calculation easier but on the
other the behavior of the longitudinal mode cannot be
determined without including the subleading $O(n^0)$ terms in the
action.

To proceed further we  self-consistently factorize
the fourth order interaction term according to~\cite{zj,mz-03}
\be
\frac{1}n \ppt\pp \pp^2\to  C^\pp_{\x=0}(t,t) \ppt \pp\,.
\ee
The model resulting from this factorization is Gaussian with a
time-dependent temperature:
\be
{\hat r(t)}=r_0+\frac{g}6  C^\pp_{\x=0}(t,t)+\frac{m^2(t)}3\,.
\ee
The response and correlation functions are ($\tau_0^{-1}=0$)
\bea
R_\q^\pp(t,s)&=& \exp\left\{ -\int_s^t \rmd t'[q^2+{\hat r(t')}]\right\}\,,\\
C_\q^\pp(t,s)&=&2\int_0^s \rmd t' R_\q^\pp(t,t')R_\q^\pp(s,t')\,,
\label{eq-expr-C}
\eea
with the self-consistency condition 
[$\hat r_\infty \equiv \hat r(t\rightarrow\infty)$]
\be
{\hat r(t)}={\hat r_\infty}+\frac{g}6
\int^\L (\rmd q) [C_\q^\pp(t,t)-C_\q^\pp(\infty,\infty)]+\frac{m^2(t)}3\,,
\label{eq-gap}
\ee
where $C_\q^\pp(\infty,\infty)=1/(q^2+\hat r_\infty)$ is the
equilibrium correlation function and $\L$ is a large-momentum cut-off.
The equation of motion~\reff{eqmotionninf} becomes
\be
\dpar_t m(t)+{\hat r(t)}m(t)=0\,,
\ee
which is solved by
\be
m(t)=m_0 \exp\left[-\int_0^t \rmd t' {\hat r(t')}\right]\,.
\ee
It is possible to cast the previous equations in a very convenient
form by introducing the function
\be
h(2\L^2 t) = \frac{1}{2} \exp\left[2 \int_0^t\!\!\rmd t' \hat r(t')\right]\,,
\label{eq-defh}
\ee
($h$ corresponds to the function $g$ usually introduced in the
literature on the dynamics of the spherical model~\cite{gl-00c,as-05}) 
in terms of which
\bea
m(t) &=& m_0 \,[2 h(2\L^2 t)]^{-1/2} ,\label{eq-m-h}\\
R_\q^\pp(t,s) &=& \rme^{-\q^2(t-s)}
\left[\frac{h(2\L^2s)}{h(2\L^2t)}\right]^{1/2}\,, \label{eq-R-h}
\eea
and the expression for $C_\q^\pp$ readily follows.
Accordingly, once we know the function $h(x)$ for a generic initial 
magnetization, the dynamics of transverse modes is completely determined.
The details of the computation of $h(x)$ are
reported in Appendix~\ref{App-h}. The final result for the critical case
($\hat r_\infty=0$), $2<d<4$ ($\e \equiv 4-d$), and in the scaling regime $t \gg (2\L^2)^{-1}$ is
\be
h(2\L^2 t \gg 1) = \frac1{u^*} \frac{\sin(\pi(d-2)/2)}{\pi
\Gamma((d-2)/{2})} (2\L^2t)^{-\e/2} 
\left[ 1 + \frac{4}{3(d-2)}m_0^2 t\right]\,,
\label{h-sol}
\ee
where $u^*=g^*N_d/(6\L^\e)$ is the fixed-point value of the four-point 
coupling (see Appendix~\ref{App-h}) and corrections to the leading scaling
behavior have been neglected.

According to Eq.~\reff{eq-m-h}, the magnetization is given by
\be
m(t) = m_0 \left[u^* \frac{\pi
\Gamma((d-2)/{2})}{2\sin(\pi (d-2)/{2})}\right]^{1/2}
(2\L^2 t)^{\e/4}\left[1 +\frac{4}{3(d-2)} m_0^2 t\right]^{-1/2}\,, 
\ee
which can be cast in the scaling form Eq.~\reff{scalm0} with 
(we recall that $a=0$, $\beta\delta/(\nu z) = (d+2)/4$~\cite{zj},
$\theta = \e/4$~\cite{jss-89}, and therefore $\kappa = 1/2$)
\be
A_m = \left[u^*\frac{\pi\Gamma((d-2)/2)}{2\sin(\pi (d-2)/2)}\right]^{1/2}
(2\L^2)^{\e/4}, \quad 
B_m = \left[\frac{4}{3(d-2)}\right]^{1/2}, 
\ee
and
\be
{\mathcal F}_M(v) = (1+v^2)^{-1/2}\,.
\ee
The time scale associated to the initial magnetization is therefore
\be
\tau_m = (B_m m_0)^{-1/\kappa} = \frac{3}{4}(d-2)m_0^{-2}\,.
\label{taumsph}
\ee
The response function follows from Eqs.~\reff{eq-R-h} and~\reff{h-sol}: 
\be
R_\q^\pp(t,s) = \rme^{-\q^2(t-s)} \left(\frac{t}{s}\right)^{\e/4}
\left[ \frac{ 1 + s/\tau_m}{
1 + t/\tau_m}\right]^{1/2}\,,
\label{Rsph}
\ee
which displays (for ${\bf q} = 0$) the expected scaling
behaviors~\reff{scalRm0}, \reff{scalR}, and~\reff{scalRT}, for
$\tau_m=\infty$, generic $\tau_m$, and $\tau_m=0$, respectively.
[We recall that $\theta_\pp = \theta'_\pp = - \beta/(\nu z) = -1/2 + \e/4$.]
In particular, comparing 
for $\tau_m = 0$  to Eq.~\reff{scalRT}, one finds $A_R^\pp = 1$,
$F_R^\pp(x)\equiv 1$.
The correlation function $C_\q^\pp(t,s)$ can be easily computed
from Eq.~\reff{Rsph} via Eq.~\reff{eq-expr-C}.
In particular, for $\q = 0$ one finds
\be
C_{\q =0}^\pp(t,s) = \frac{4}{d}
s\left(\frac{t}{s}\right)^{\e/4} \left[ \frac{ 1 + s/\tau_m}{
1 + t/\tau_m}\right]^{1/2}\left[1+\frac{2}{d-2} \frac{1}{1+s/\tau_m} \right]\,,
\label{Csph}
\ee
which displays also the expected scaling
behaviors~\reff{scalCm0}, \reff{scalC}, and~\reff{scalCT}, for
$\tau_m=\infty$, generic $\tau_m$, and $\tau_m=0$, respectively.
In particular Eq.~\reff{scalRT} holds with $A_C^\pp = 4/d$ and
$F_C^\pp(x)\equiv 1$.

Equations~\reff{Rsph} and~\reff{Csph} allow the computation of the FDR:
\be
{\mathcal X}_{\q=0}^\pp(u) = \frac{d}{d+2}
\left[ 1 + \frac{4}{d^2-4} \frac{1}{(1+u)^2}\right]^{-1}\,,
\label{Xuninf}
\ee
where  $u \equiv s/\tau_m$. For $m_0=0$ (\ie\ $s \ll \tau_m \propto
m_0^{-2}$) we recover the well-known result $X^\infty=1-2/d$
\cite{gl-00c,cg-04}, whereas for $m_0\neq 0$ 
the long-time asymptotic value for 
$s\gg \tau_m$ is given by
\be
X^\infty_\pp(n=\infty)=\frac{d}{d+2}\,.
\label{Xinfninf}
\ee
Close to $d=2$ this expression becomes
\be
X^\infty_\pp(n=\infty) = \frac{1}{2} + \frac{1}{8} \tilde\epsilon + O(\tilde\epsilon^2)
\ee
where $\tilde \epsilon = d-2$, in agreement with the limit
$n\rightarrow\infty$ of the result of Ref.~\cite{ft-05}
(see Eq.~(18) therein): 
\be
X^\infty = \frac{1}{2} + \frac{1}{8}\frac{n-1}{n-2} \tilde\epsilon + O(\tilde\epsilon^2)\,.
\label{XinfFT}
\ee
In passing we note that Eq.~\reff{Xinfninf} 
coincides with the FDR derived in Ref.~\cite{as-05} for local spin 
observables in the spherical model (see Eq.~(8.12) therein).

A plot of Eq.~\reff{Xinfninf} is provided as a dashed line in
Fig.~\ref{Xinfs} (right panel) together with the result of the
expansion~\reff{XinfFT} for $n=3$ and $n=\infty$ (denoted by
$n=3 [\tilde\epsilon]$ and $n=\infty [\tilde\epsilon]$, respectively).
As in the Gaussian case, ${\mathcal X}_{\q=0}^\pp(u)$ can be used to describe
the crossover from magnetized to non-magnetized initial state and the
scaling function resembles closely  Eq.~\reff{Xcrossu}, --- derived for
the Gaussian model and corresponding to the case $d>4$ ---
the only difference being the values at $u=0$ and for $u\rightarrow
\infty$ and the actual expression for $\tau_m$ as a function of
$m_0$. In Fig.~\ref{Xcombi}(a) we compare the exact scaling
function for $d=3$ (see Eq.~\reff{Xuninf}) to the Gaussian one
(Eq.~\reff{Xcrossu}, formally $d=4$). 
In both cases ${\cal X}_{{\bf q}=0}^\pp$ increases monotonically
from the value $1-2/d$ for $u=0$ to $d/(d+2)$ for
$u\rightarrow\infty$.

\begin{figure}[tb]
\centerline{\epsfig{width=9truecm,file=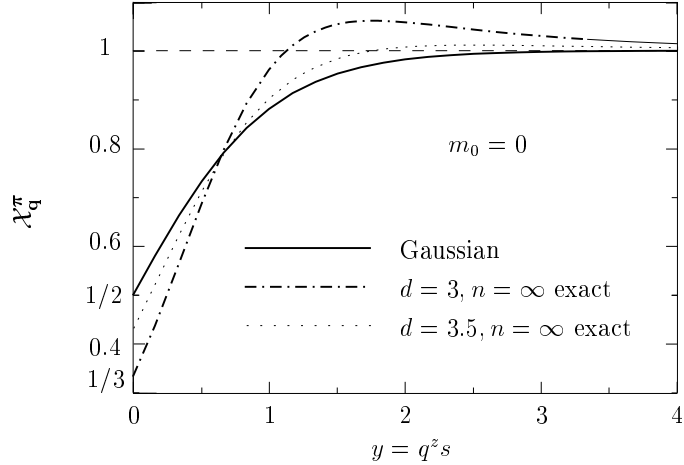}}
\caption{Non-monotonic behavior of ${\mathcal X}_\q^\pp$ as a function
of $y=q^z s$ and different values of $2<d<4$ for the exactly solvable
case $n\to\infty$ of the $O(n)$ model 
relaxing from a {\it disordered} initial state ($m_0=0$). 
The Gaussian result corresponds to $d > 4$.}
\label{nonmon}
\end{figure}

As in the case of the Gaussian approximation (see Sec.~\ref{sec-Gaux}), 
we consider the momentum-dependence of the FDR.
As a function of the two scaling variables $y=q^z s = q^2 s$
and $u = s/\tau_m$ (where $\tau_m$ is given in Eq.~\reff{taumsph})
one finds, from Eqs.~\reff{Rsph} and~\reff{eq-expr-C}:
\be
{\mathcal X}_\q^\pp = \frac{1}{2} 
\left[ 1 - \left(\frac{d-2}{4} + y - \frac{1}{2}\frac{1}{1+u}\right) 
\int_0^1\!\!\rmd x\, (1-x)^{(d-4)/2} \left(1 - \frac{u}{1+u} x \right)
\rme^{-2 x y}\right]^{-1}
\label{Xuy}
\ee
from which one recovers Eq.~\reff{Xuninf} for $y=0$ and the Gaussian result~\reff{XyGaux} for $d=4$ and $u
\rightarrow \infty$, as expected.
Taking into account that
$
\int_0^1\!\rmd x\, (1-x)^{-\e/2} \rme^{- 2 y x} = 1/(2 y) +
\e/(8 y^2) + \e (2+\e)/(32 y^3) + O(y^{-4})
$
one easily determines the behavior of ${\mathcal X}_\q^\pp$ for $y
\gg 1$:
\be
{\mathcal X}_\q^\pp = 1 - \frac{1}{4 y^2} \left[ \left(
\frac{u}{1+u}\right)^2 - \frac{\e}{2}\right] + O(y^{-3}) \,.
\label{Xto1}
\ee
Accordingly, for $y = q^2 s \gg 1$ the FDR converges to the
asymptotic value $1$, as heuristically expected on the basis of the
fact that the mode actually responsible for the aging behavior is the
homogeneous one $\q = 0$. We note also that this asymptotic value is
approached from above for $0\le u < 1/(\sqrt{2/\e}-1)$ and from below
for $u > 1/(\sqrt{2/\e}-1)$. 
Given that ${\mathcal X}_\q^\pp < 1$ for $y=0$,
${\mathcal X}_\q^\pp$ has a non-monotonic 
behavior as a function of $y$ for fixed $0\le u <  1/(\sqrt{2/\e}-1)$
and in particular for $u=0$, corresponding to an initially disordered
configuration. This non-monotonicity disappears in the Gaussian limit
$d\rightarrow 4$ and was already noticed at the first order in the
$\e$-expansion for the relaxation of the $O(n)$ model from a
disordered state
(see Fig.~1 in Ref.~\cite{g-05} and
Ref.~\cite{cg-02a1} for details). In fact, the expansion in $\e=4-d$ of 
Eq.~\reff{Xuy} for $u=0$ agrees with the limit $n\rightarrow \infty$
of the results presented in Ref.~\cite{cg-02a1} (cp. Eq.~(3.14)
therein). 
In Fig.~\ref{nonmon} we provide a plot of
Eq.~\reff{Xuy}, as a function of $y=q^2s$ for $m_0=0$ and $d=3$,
$3.5$, and $4$, the latter corresponding to the Gaussian model.
The non-monotonic behavior is still quite pronounced for $d=3$ and
the overshoot decreases upon increasing the dimensionality $d$
(cf. the dotted line corresponding to the unphysical dimension
$d=3.5$, reported for comparison), while shifting towards larger values
of $y$. 
In the opposite limit, \ie\ $u\rightarrow\infty$
(corresponding to $\tau_m=0$), a monotonic increase of
${\mathcal X}_\q^\pp$ as a function of $y=q^2s$ is expected. In
Fig.~\ref{Xcombi}(b) we compare the behavior predicted in this case
by Eq.~\reff{Xuy} (with $u\rightarrow \infty$) for $d=3$,
to the Gaussian result~\reff{XyGaux} (formally, $d=4$). We also indicate the
limiting curve for $d\rightarrow 2$. In these cases ${\mathcal
X}_\q^\pp$ increases monotonically from its initial value, given by
$d/(d+2)$, to the asymptotic value $1$, which is approached according
to Eq.~\reff{Xto1}.

\section{One-loop fluctuation-dissipation ratios}
\label{sec-onel}

In the previous sections we have discussed the non-equilibrium
behavior of models whose action is essentially quadratic in the
fields (either because non-Gaussian terms were neglected or turned
into effective Gaussian ones). In this section we account for the
effects of non-Gaussian fluctuations within the perturbation theory
up to one-loop (i.e., $O(\e)$ in the
$\e$-expansion where $\e = 4 - d$).
In particular we compute the
longitudinal and transverse correlation and response functions for a
quench from a state with initial magnetization $m_0\to \infty$ to
the critical point. As we have argued from general scaling
arguments, a finite $m_0$ gives only corrections to the leading
long-time behavior. The first step in this calculation is the
solution of the one-loop equation of motion, presented in the next
subsection. Then we calculate the connected (longitudinal and
transverse) correlation and response
functions (and the associated FDRs) in the following subsections,
reporting the relevant details in Appendix~\ref{app-conti}.

For future reference we recall that the standard static exponents
$\beta$, $\delta$, $\nu$ of the universality class we are presently
interested in are such that $\beta/\nu = (d-2+\eta)/2$,
$\delta = (d+2-\eta)/(d-2+\eta)$  with $\eta = O(\e^2)$ and
$z=O(\e^2)$ (see, e.g.,~\cite{zj}),
and therefore [cp. Eqs.~\reff{conclL} and~\reff{conclT}]
\be
a = O(\e^2)\,,\quad
\theta_\s = -\frac{\beta\delta}{\nu z} = -\frac{3}{2} + \frac{\e}{4} +
O(\e^2) \,,\ \mbox{and}\quad \theta_\pp = -\frac{\beta}{\nu z} = -
\frac{1}{2} + \frac{\e}{4} + O(\e^2)\,.
\label{exp1loop}
\ee

\subsection{The equation of motion and its solution}

At one-loop level the tadpole contributions to the equation
of motion~\reff{eqmo} are
\bea
I_\s(t) &=&\int (\rmd q) C^{\s,0}_{\bf q} (t,t)= 2 N_d r_\s  t^{1-d/2}\,,\label{Is}\\
I_\pp(t)&=&\int (\rmd q) C^{\pp,0}(t,t)_{\bf q}= 2 N_d r_\pp t^{1-d/2}\,,\label{Ip}
\eea 
where $N_d=2/[(4\pi)^{d/2} \Gamma(d/2)]$, and
\bea
r_\s &=& \frac{96}{(8\pi)^{d/2} (8-d)(6-d)(4-d)(2-d)N_d}\,,\\
r_\pp&=& \frac{4}{(8\pi)^{d/2} (4-d)(2-d)N_d}\,.
\eea
Accordingly, the one-loop equation of motion is [see Eq.~\reff{eqmo}]
\bea
0&=&\dpar_t m +\frac{1}{3} m^3+ g_0 N_d\left(r_\s+\frac{n-1}{3}r_\pp
\right) t^{1-d/2} m\equiv + O(g_0^2)\\&\equiv&
\dpar_t m +\frac{1}{3} m^3+\gt_0 r_d t^{1-d/2} m + O(\gt_0^2)\,,
\eea
where $\gt_0 \equiv N_d g_0$ and
\be
r_d \equiv r_\s + \frac{n-1}{3} r_\pp \;,
\label{rd}
\ee
which is solved by
\bea
m(t)&=&\sqrt{\frac{3}{2t}}\left[1- \frac{\gt_0}{3-d/2} r_d t^{\e/2}\right]
+ O(\gt_0^2)\,.
\label{m1loop0}
\eea
Expanding in $\e=4-d$, one obtains
\bea
m(t)&=&
 \left(1-\gt_0 \frac{c_{-1}}{\e}\right) \sqrt{\frac{3}{2t}}
\left[1-\gt_0 \left(\frac{c_{-1}}{2} \ln t+c_0-\frac{c_{-1}}{2}\right)
+ O(\e^2,\e\gt_0,\gt_0^2)\right] \,,
\label{m1loop}
\eea
where $c_{-1}=-({n+8})/{12}$ and $c_0=9/16-(n+8)(\ln2+\gamma_E)/{24}$
are the coefficients of the expansion of $r_d = c_{-1}/\e + c_0 + O(\e)$.
The dimensional pole $\sim 1/\e$ can be cancelled out by a proper
renormalization of the bare magnetization (see, e.g.,
Ref.~\cite{zj,bj-76,bej-79,cgk-06} for details). Once the renormalized $m(t)$
is expressed in terms of $\gt = \gt_0 + O(\gt_0^2)$, 
its scaling behavior at the RG fixed
point $\gt^* = 6\e/(n+8) + O(\e^2)$
clearly shows up: $m(t)\sim
t^{-\varsigma} + O(\e^2)$ with
$\varsigma =1/2(1+c_{-1}\gt^*) + O(\e^2) = \beta/(\nu z)$.

\subsection{The Longitudinal response function}

\begin{figure}[tb]
\centerline{\epsfig{width=9truecm,file=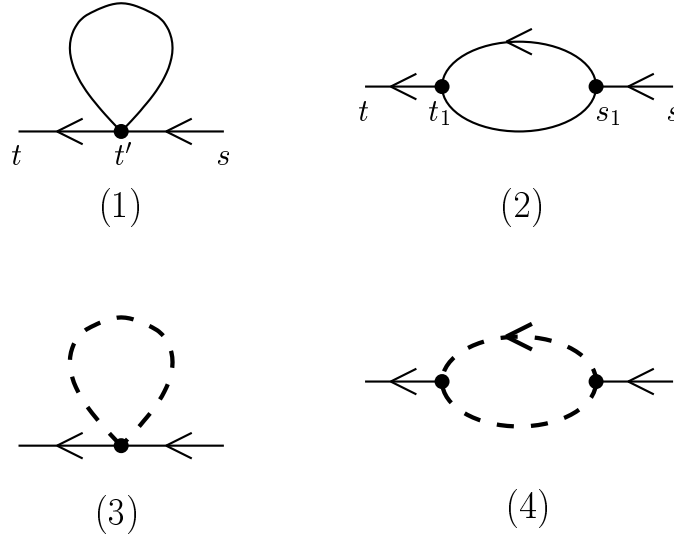}}
\caption{One-loop diagrams contributing to the longitudinal response function
$R_\q^\s(t,s)$. The corresponding expressions (reported in
Appendix~\ref{appRL}) are indicated by $I_i(t,s)$, $i=1,\ldots 4$.}
\label{diagRL}
\end{figure}

At one-loop order the expression for the response
function gets modified because of the one-loop term
contributing to the magnetization $m(t)$ in Eq.~\reff{m1loop}, and therefore
\bea
R^{\s,0}_{\q}(t,s) &=& \exp\left[ -q^2(t-s) - \int_s^t \rmd t' m^2(t')\right]\nonumber\\
&=& \left(\frac{s}{t}\right)^{3/2} \left[ 1 + \gt_0 \frac{3}{1+\e/2} r_d
\frac{t^{\e/2}-s^{\e/2}}{\e/2} + O(\gt_0^2)\right]\rme^{-q^2(t-s)}
\eea
In addition, the
interaction vertices of  ${\cal L}_1$ [see Eq.~\reff{L1}] yield four
contributions which are depicted in Fig.~\ref{diagRL}.
In terms of those diagrams the response function is
given by
\be
R^\s_\q(t,s)=R^{\s,0}_\q(t,s)-\frac{g_0}2I_1-g_0\frac{n-1}{6} I_3+
g_0 I_2+g_0\frac{n-1}{9} I_4 + O(g_0^2)\,.
\ee
The relevant zero-momentum integrals $I_i$, $i=1,\ldots,4$,  
have been calculated in Appendix~\ref{appRL}.
Collecting the corresponding expressions we obtain, in terms of the
renormalized coupling $\gt$ and of the ratio $x\equiv s/t$ ($0\le x
\le 1$),
\be
R^\s_{\q=0}(t,s)= x^{3/2}\left[1+\tilde{g}\left\{
-\frac{n+8}{24}\ln x+\left(\frac{\pi^2}4-\frac{61+2n}{24}+f^\s_R(x)
\right)\right]\right\} + O(\gt^2,\e\gt)\,,
\label{R1loop}
\ee
where
\be
f^\s_R(x) = \frac{26+n}{12}\left[1+\ln (1-x)\left(\frac1x-1\right) \right]
+\frac38 x-\frac32\Li_2(x)\,,
\ee 
with $f^\s_R(0) = 0$. 
At the fixed point $\gt^*=6/(n+8) \e + O(\e^2)$,
Eq.~\reff{R1loop} can be cast in 
the expected scaling form~\reff{scalRL} with
\bea
A_R^\s&=&1+\gt^*\left[\frac{\pi^2}4-\frac{61+2n}{24}\right]+O(\e^2)\,,\label{ARS}\\
F^\s_R(x)&=&1+\gt^* f^\s_R(x) +O(\e^2)\,.
\eea

\subsection{Longitudinal Connected Correlation function}

\begin{figure}[tb]
\centerline{\epsfig{width=12truecm,file=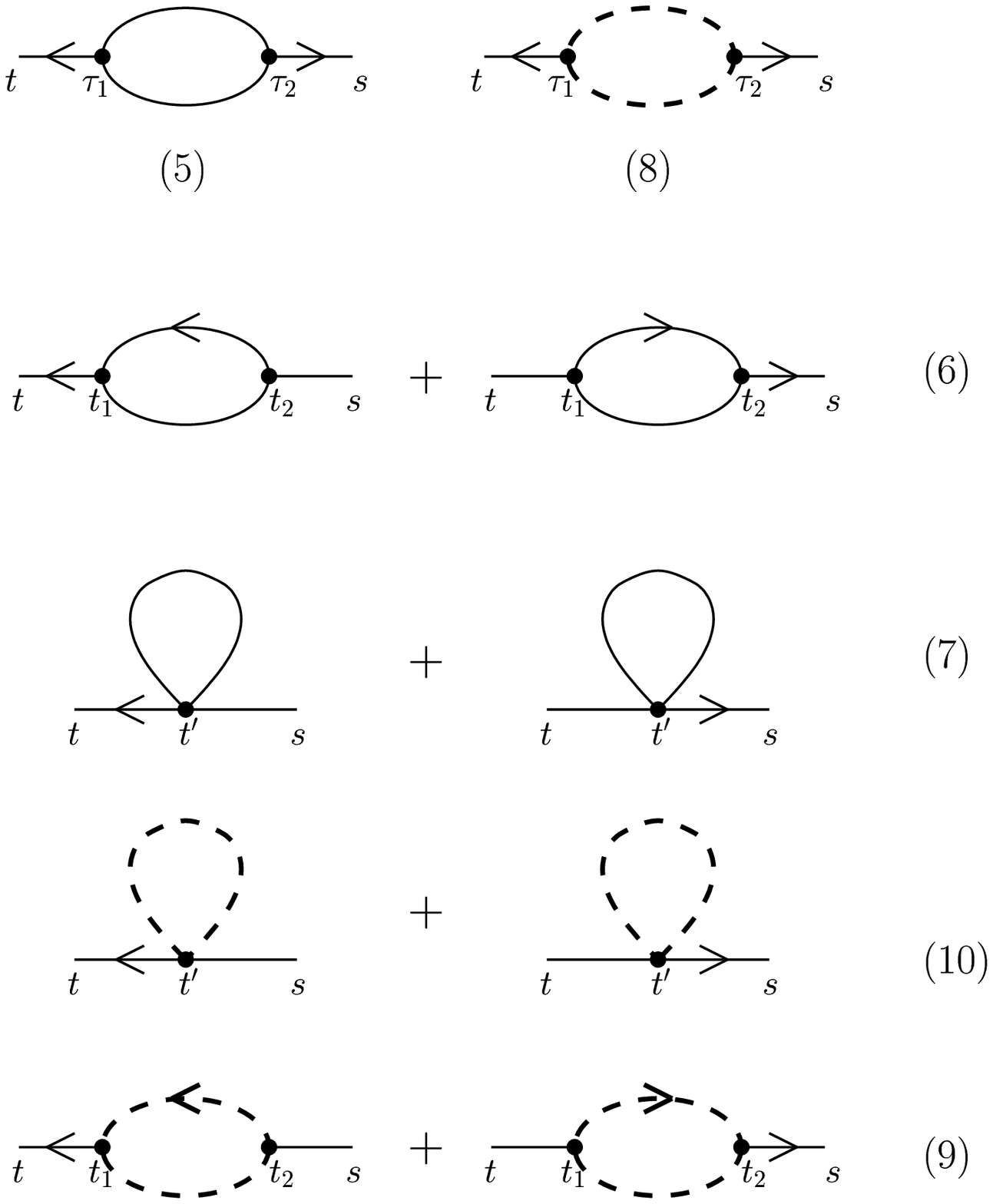}}
\caption{One-loop diagrams contributing to the longitudinal correlation
function
$C_\q^\s(t,s)$. The corresponding expressions are indicated by
$I_i(t,s)$, $i=5,\ldots 10$ and are computed in Appendix~\ref{appCL}}
\label{diagCL}
\end{figure}

As in the case of $R_{\bf q}^\s$, the longitudinal correlation
function gets a one-loop correction $C^{\s,0}_\q(t,s)$ due the one-loop term
in the magnetization $m(t)$ [see Eq.~\reff{m1loop}].
This correction has been evaluated in Ref.~\cite{cgk-06},
Eq.~(B.29). In making use of that result we have to keep in mind that 
the factor $r_d$ in  Ref.~\cite{cgk-06} accounts only for the
contribution of longitudinal modes (i.e., $r_d$ would be $r_\s$ in the current
notation) whereas here it contains also the contribution of 
transverse modes [see Eq.~\reff{rd}]. The additional interaction
vertices contribute via the Feynman diagrams depicted in
Fig.~\ref{diagCL}, in terms of which
\be
C^\s_\q(t,s) = C^{\s,0}_\q(t,s)+\frac{g_0}{2} I_5+g_0\frac{n-1}{18} I_8
+g_0 I_6+g_0\frac{n-1}{9} I_9
-\frac{g_0}{2} I_7 - g_0\frac{n-1}{6} I_{10}+O(g_0^2)\,.
\label{CsumS}
\ee
Using the results reported in Appendix~\ref{appCL} we obtain the zero-momentum
longitudinal correlation function ($x\equiv s/t$)
\be
C^\s_{\q=0}(t,s)=\frac{1}{2}s x^{3/2}
\left[1+\gt \left(-\frac{n+8}{24}\ln x +
\frac{37 \pi ^2}{160}-\frac{25 n}{432}-\frac{18913}{8640}
\right)+\gt f_C^\s(s/t) 
+O(\gt^2,\e\gt)
\right] \,,
\label{CresultS}
\ee
which has, at the RG fixed point, the expected scaling form with the
correct exponents and
\bea
2 A_C^\s&=&1+\gt^*
\left(\frac{37 \pi ^2}{160}-\frac{25 n}{432}-\frac{18913}{8640}\right)+O(\e^2)\,,\label{ACS}\\
F_C^\s(x)&=&1+\gt^*f_C^\s(x)+O(\e^2)\,,
\eea
where $f_C^\s(x)$ is given by Eq.~\reff{fcs}.

\subsection{Longitudinal FDR}

It is now easy to compute the longitudinal FDR from Eqs.~\reff{Xinfamprat},
\reff{conclL}, \reff{ARS}, and~\reff{ACS}
\be
X^\infty_{\s}=\frac{A_R^\s}{A^\s_C(1-\theta_\s)}=
\frac{4}{5}-\frac{1895 + 76 n - 162 \pi^2}{1800 (8 + n)}\e +O(\e^2)\,,
\label{Xinfeps}
\ee
which for $n=1$ reproduces the result of 
Ref.~\cite{cgk-06} for the Ising model.
Close to $d=4$, $X^\infty_{\s}$ ($n\geq 2$) decreases as 
the dimensionality decreases. 
A monotonic behavior has to be expected down to the lower critical 
dimension $d=2$, where it is known that $X^\infty_{\s}=1/2$ \cite{ft-05}.
In the limit $n\to\infty$ the FDR becomes
\be
X^\infty_{\s}(n=\infty)=\frac{4}{5}-\frac{19}{450}\e+O(\e^2)\,,
\label{XCG}
\ee
which differs from the $\e$-expansion of the result reported in Ref.~\cite{as-05} (see Eq.~(8.116) therein) for the
spherical model:
\be
X^\infty({\rm sph}) = \frac{4}{5} + \left( \frac{2 \ln 2}{15} - \frac{53}{900}
- \frac{\pi^2}{120}\right) \e + O(\e^2) \,.
\label{XAS}
\ee
\begin{figure}[tb]
\begin{minipage}{0.48\textwidth}
\centerline{\epsfig{width=\textwidth,file=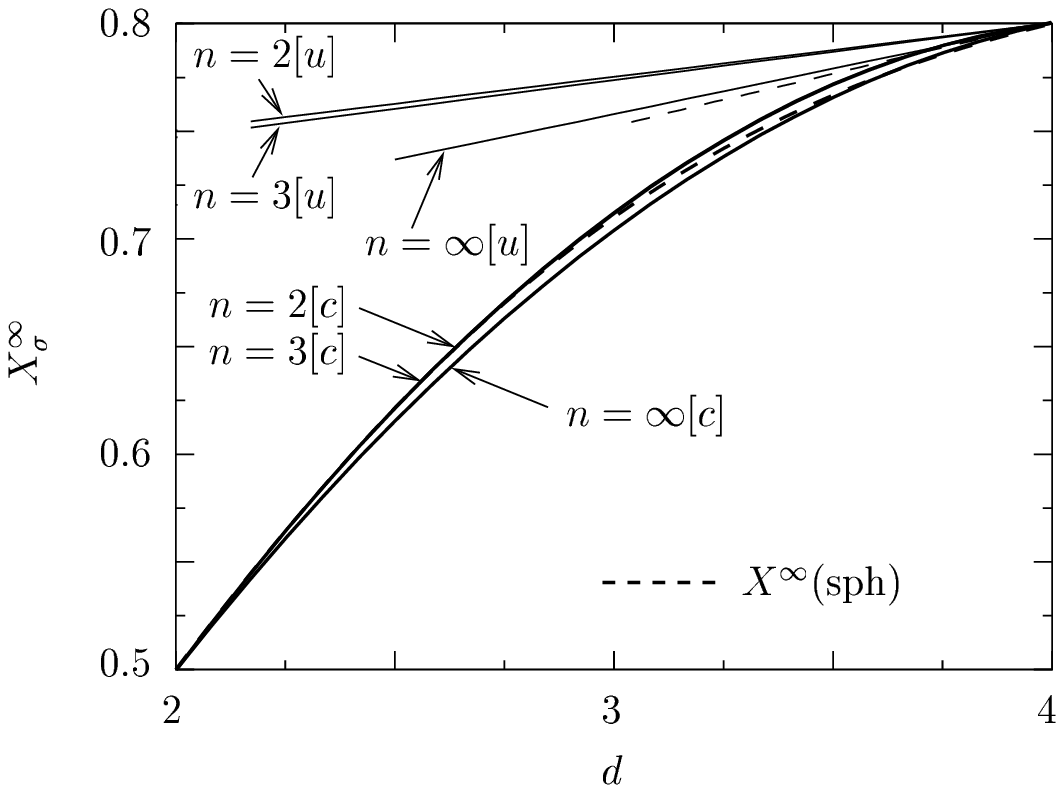}}
\end{minipage}
\hspace{2mm}
\begin{minipage}{0.48\textwidth}
\centerline{\epsfig{width=\textwidth,file=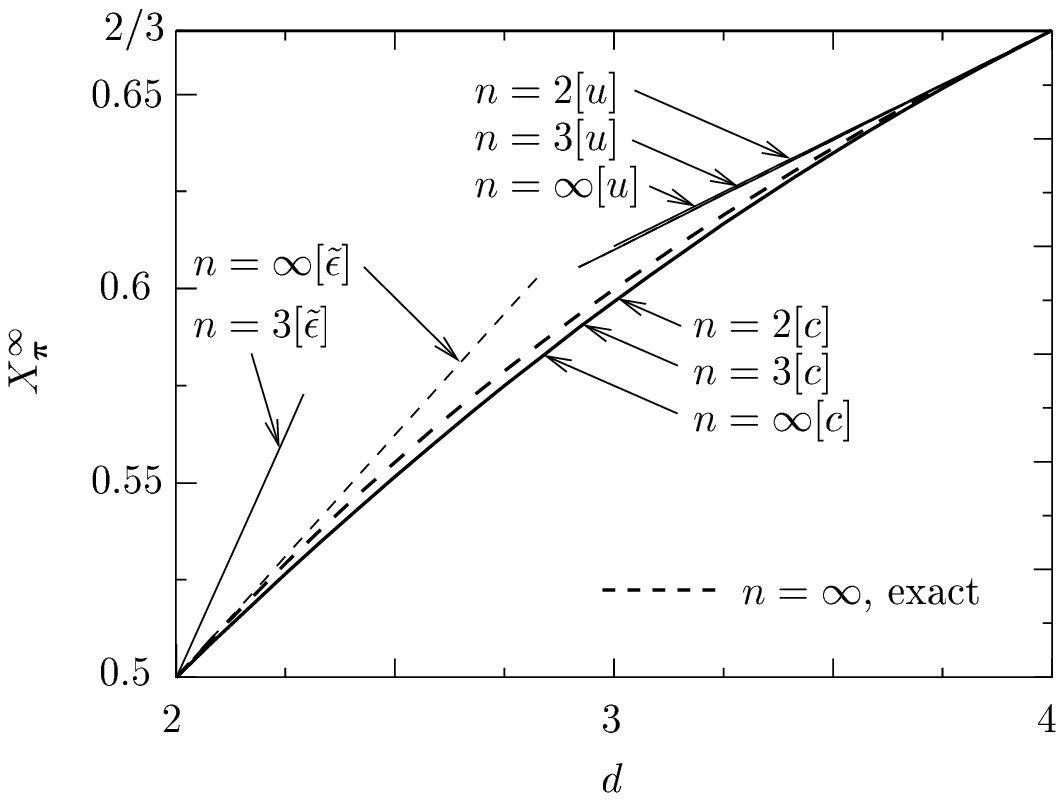}}
\end{minipage}
\caption{Left: Longitudinal FDR $X^\infty_\s$ as a function of the
dimensionality $d$.
The unconstrained ([u], see Eq.~\reff{Xinfeps}) and constrained ([c], see Eq.~\reff{XinfLc}) 
estimates for $n=2,3$ and $\infty$
are shown together with the result $X^\infty({\rm sph})$ of
Ref.~\protect\cite{as-05} for the spherical model (dashed line). 
The dashed straight line indicates the expansion of $X^\infty({\rm
sph})$ close to $d=4$, which does not coincide with the limit $n\to
\infty$ of Eq.~\reff{Xinfeps}.
Right: Transverse FDR $X^\infty_\pp$ as a function of the
dimensionality $d$. The unconstrained ([u], see Eq.~\reff{Xinfpexpr}) and constrained ([c], see Eq.~\reff{Xsc}) 
estimates for $n=2,3$ and $\infty$
are shown together with the exact result~\reff{Xinfninf} for
$n=\infty$ (dashed line). For comparison we show also the results of
the $\tilde\e$-expansion ($\tilde\e = d-2$) of
Ref.~\protect\cite{ft-05}, 
Eq.~\reff{XinfFT}, for $n=3$ and $n=\infty$.}
\label{Xinfs}
\end{figure}
For finite $n$, a first rough numerical estimate of 
$X^\infty_{\s}$ in the 
interesting and non-trivial physical dimension $d=3$ can be obtained by 
evaluating equation~\reff{Xinfeps} for $\e=1$. 
For instance we have ($[u]$ denotes 
unconstrained estimates, to distinguish them from the constrained ones
introduced below)
\bea
X^\infty_\s[u](n=2)&=& 0.775\,,\\ 
X^\infty_\s[u](n=3)&=& 0.774 \,.
\eea
However, the reliability of such results can be questioned, both 
because they are based on a one-loop approximation and mainly because 
for $\e=2$ one would find $X^\infty_\s[u]\sim 0.75$, which is quite far 
from the exact result $1/2$. 
To obtain more reliable estimates without computing the 
$O(\e^2)$  contribution (which seems to be a difficult task, requiring
the calculation of almost fifty Feynman diagrams) one can constrain
$[c]$ the $O(\e)$ result to assume the exact known value at $d=2$, as
usually done for this kind  of expressions (see, \eg\ \cite{cg-04,PV-r}).
Assuming a smooth behavior in $\e$ up to $\e=2$ one gets
\be
X^\infty_\s[c]=\frac12  + \left(1- \frac\e2\right) 
\left[\frac3{10} - \left(\frac{1895 + 76 n - 162 \pi^2}{1800(8 + n)} - \frac3{20}\right)\e\right]\,,
\label{XinfLc}
\ee
which has the same $\e$-expansion as Eq.~\reff{Xinfeps}, assumes the
exact value in $d=2$ and therefore it is expected to
converge more rapidly to the correct three-dimensional result.
The estimate~\reff{XinfLc} yields, for $\e=1$,
\bea
X^\infty_\s[c](n=2)&=& 0.713\,,\\ 
X^\infty_\s[c](n=3)&=& 0.712 \,,
\eea
which are considerably smaller than the corresponding unconstrained estimates.
A more robust field-theoretical estimate would require the knowledge
of higher-order terms in the $\e$-expansion.
In Fig.~\ref{Xinfs} we compare the unconstrained $[u]$ and constrained
$[c]$ estimates for $n=2,3,\infty$ 
as functions of the dimensionality $d$ ranging between the
lower and the upper critical dimensions of the model ($2$
and $4$, respectively). 
For comparison we report the exact result $X^\infty({\rm
sph})$ for the spherical
model (dashed line), given by Eq.~(8.113) in Ref.~\cite{as-05}
(cp.~also Fig.~4 therein). As already pointed out, our
result~\reff{XCG} close to $d=4$ (see the straight line $n=\infty [u]$
in Fig.~\ref{Xinfs}) differs from the corresponding one~\reff{XAS} 
obtained in Ref.~\cite{as-05} (represented as a straight dashed line in
Fig.~\ref{Xinfs}). On the other hand, the numerical discrepancy between
these two expressions is quite small, resulting in a 
constrained estimate $X^\infty_\s[c](n=\infty)$ which differs from
$X^\infty({\rm sph})$ by at most $3\%$.   

\begin{figure}[tb]
\centerline{\epsfig{width=9truecm,file=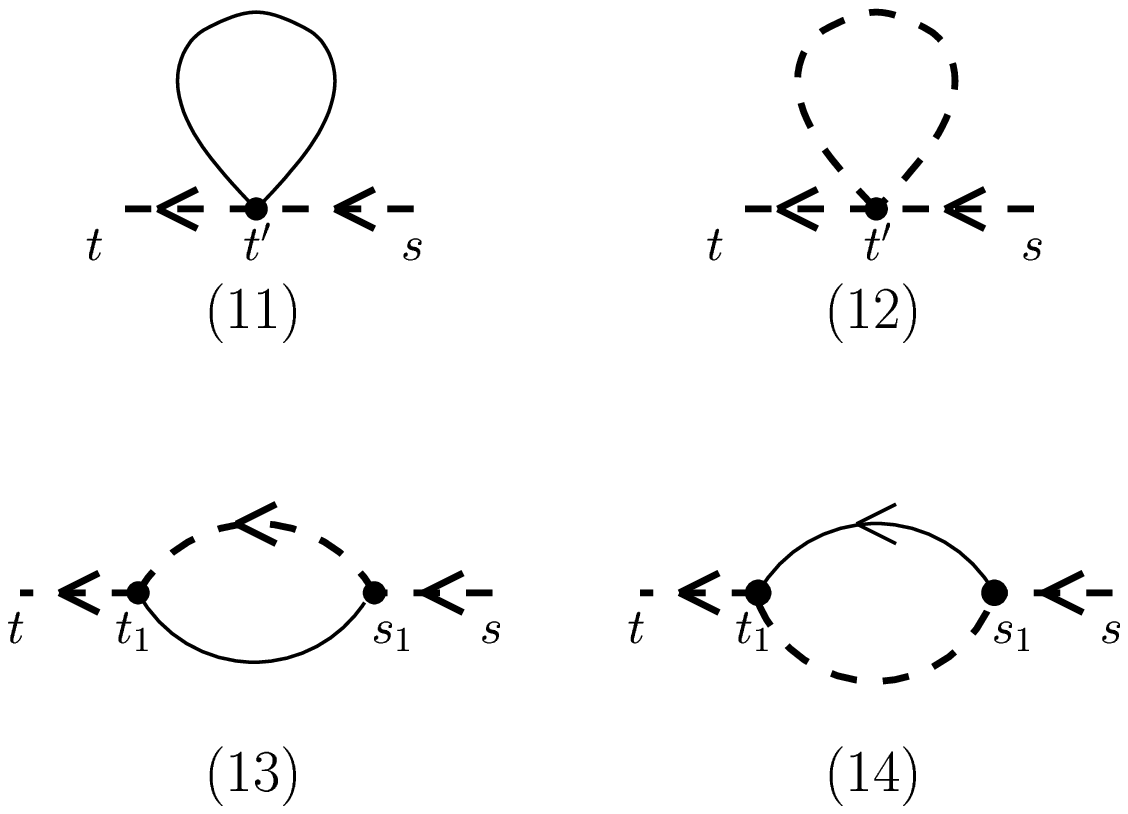}}
\caption{One-loop diagrams contributing to the transverse response function
$R_\q^\pp(t,s)$. The corresponding expressions, indicated by
$I_i(t,s)$, $i=11,\ldots,14$, are calculated in Appendix~\ref{appRT}.}
\label{diagRT}
\end{figure}

\subsection{Transverse response function}

The transverse response function, like the longitudinal one,  gets 
modified because of the one-loop contribution to $m(t)$:
\bea
R^{\pp,0}_{{\bf q}=0}(t,s) &=& \rme^{-q^2(t-s) - \frac13\int_s^t \rmd t' m^2(t')}\nonumber\\
&=& \left(\frac{s}{t}\right)^{1/2} \left[ 1 + \gt_0 \frac1{1+\e/2} r_d
\frac{t^{\e/2}-s^{\e/2}}{\e/2} + O(\gt_0^2)\right] \rme^{-q^2(t-s)}\,.
\label{Rppm1loop}
\eea
In addition, the interaction vertices in ${\cal L}_1$ [see
Eq.~\reff{S123}] yield the diagrammatic contributions depicted in 
Fig.~\ref{diagRT}, in terms of which the transverse response function reads
\be
R^\pp_\q(t,s)=R^{\pp,0}_\q(t,s)-\frac{g_0}{6} I_{11}-g_0\frac{n+1}{6} I_{12}+
\frac{g_0}9 I_{13}+\frac{g_0}9 I_{14} + O(g_0^2)\,.
\ee
Using the expressions of $I_i$, $i=11,\ldots,14$\ reported in 
Appendix~\ref{appRT} we find for
${\bf q} = 0$ ($x\equiv s/t$)
\bea
R^\pp_{\q=0}(t,s)&=& x^{1/2}\left[1+\gt\left(-\frac{n+8}{24}\ln x-\frac1{24}+
f_R^\pp(x) + O(\gt^2,\e\gt)
\right)\right]\,,
\label{RP1loop}
\eea
where 
\be
f_R^\pp(x) = \frac{ \ln (1-x) + x}{4 x} -
\frac5{24} x -\frac13 \ln (1-x)+\frac1{12} x\ln(1-x)
\ee
($f_R^\pp(0) = 0$).
Equation~\reff{RP1loop} displays, at the RG fixed point $\gt = \gt^*$,
the correct scaling behavior [see Eq.~\reff{scalRT}] with the proper
exponents [see Eq.~\reff{exp1loop}] and
\bea
A_R^\pp&=&1-\frac{\gt^*}{24}+O(\e^2)\,,\label{ARP}\\
F_R^\pp(x)&=&1+\gt^* f_R^\pp(x) 
+O(\e^2)\,.
\eea

\subsection{Transverse correlation}

\begin{figure}[tb]
\centerline{\epsfig{width=12truecm,file=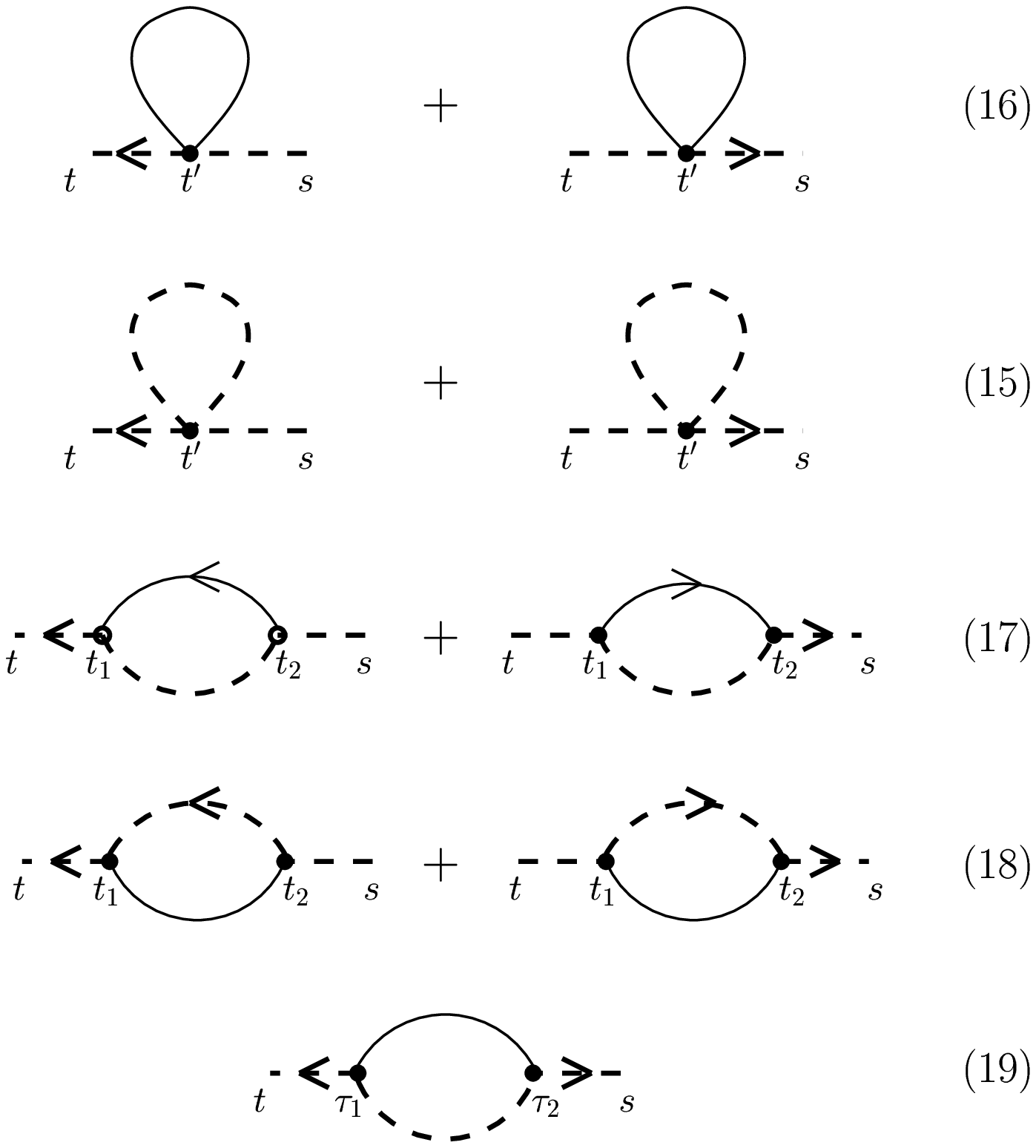}}
\caption{One-loop diagrams contributing to the transverse correlation
function
$C_\q^\pp(t,s)$. The corresponding expressions, indicated by
$I_i(t,s)$, $i=15,\ldots,19$, are reported in Appendix~\ref{appCT}.}
\label{diagCT}
\end{figure}

As in the case of $C_{\bf q}^\s$, the transverse correlation
function gets a one-loop correction $C^{\pp,0}_\q(t,s)$ due the one-loop term
in the magnetization $m(t)$ [see Eq.~\reff{m1loop}]. This correction
can be computed from Eqs.~\reff{Cgaux} and~\reff{Rppm1loop}, finding 
\bea
C^\pp_{\q = 0}(t,s) &=& 2 \int_0^s\!\rmd t'\, R^{\pp,0}_{{\bf
q}=0}(t,t') R^{\pp,0}_{{\bf q}=0}(s,t') \nonumber\\
&=& 
s\, x^{1/2} \left\{ 1 + \gt_0 \frac{r_d}{1+\e/2} s^{\e/2} \left[
\frac{x^{-\e/2} -1}{\e/2} + \frac{2}{2+\e/2}\right]\right\} + O(\gt_0^2)\;.
\label{Cppm1loop}
\eea
The additional vertices contribute via the Feynman diagrams depicted
in Fig.~\ref{diagCT}, according to
\be
C^\pp_\q(t,s) = C^{\pp,0}_\q(t,s)-
g_0 \frac{n+1}{6} I_{15} -\frac{g_0}6 I_{16}
+\frac{g_0}9 I_{17}+\frac{g_0}{9} I_{18}+
\frac{g_0}{9} I_{19}+O(g_0^2)\,.
\label{CsumP}
\ee
Using the results reported in Appendix~\ref{appCT} for $I_i$,
$i=15,\ldots,19$,
we obtain the zero-momentum transverse correlation function
\be
C^\pp(t,s)_{\q=0}=s\left(\frac{s}{t}\right)^{1/2}
\left[1+\gt \left(-\frac{n+8}{24}\ln x +
\frac{127}{432}+\frac{n}{24}+ f^\pp_C(s/t)\right) +O(\gt^2,\e\gt)\right] \,,
\label{CresultP}
\ee
which displays the expected scaling behavior [cp. Eq.~\reff{scalCT}]
with the proper exponents [see Eq.~\reff{exp1loop}] and
\bea
A_C^\pp&=&1+\gt^*\left(\frac{127}{432}+\frac{n}{24}\right)+O(\e^2)\,,\label{ACP}\\
F^\pp_C(x)&=&1+\gt^* f^\pp_C(x)+O(\e^2)\,,
\eea
where $ f^\pp_C(x)$ is given by Eq.~\reff{fcp}.
%
%
%
%

\subsection{The transverse FDR}

The expressions of the transverse response and correlation functions
allow the determination of the FDR according to Eq.~\reff{Xinfamprat} 
[see also
Eqs.~\reff{ARP}, \reff{ACP}, and~\reff{conclT}]
\be
X^\infty_{\pp}=\frac{A_R^\pp}{A^\pp_C(1-\theta_\pp)}=
\frac{2}{3}-\frac{49 + 6 n}{108 (8 + n)}\e +O(\e^2)\,,
\label{Xinfpexpr}
\ee
which is a decreasing function of $\e$, a feature that we expect
to persist down to the lower critical dimension ($d=2$ for the case of
interest $n\geq2$).
Note that the coefficient of the correction of $O(\e)$, \ie\ $1/18 +
1/[108 (8 + n)]$, changes less than $2\%$ with $n\ge 2$ and
therefore $X^\infty_{\pp}$, at least close to $d=4$, is practically 
independent of the actual number $n-1$ of transverse components.
In the limit $n\to\infty$, $X^\infty_{\pp}$ reduces to
\be
X^\infty_{\pp}(n=\infty)=\frac{2}{3}-\frac{1}{18}\e+O(\e^2)\,,
\ee
in agreement with the $\e$-expansion of the exact result 
$X^\infty_{\pp}(n=\infty)=d/(d+2)$, obtained in Sec. \ref{Secninf}.
As for finite $n$, we proceed as in the case of the longitudinal FDR, first 
by giving unconstrained estimates (obtained by setting $\e$ to the
proper value in Eq.~\reff{Xinfpexpr} and neglecting $O(\e^2)$
corrections) and then improving them by
taking advantage of the known value for $d = 2$. 
The direct estimates for the physical dimension $d=3$ are 
(they differ on the fourth digit)
\be
X^\infty_\pp[u](n=2,3)= 0.610\,. 
\ee
These unconstrained $[u]$ estimates
are expected to work better than in 
the longitudinal case, because for $\e=2$ (i.e., $d=2$) 
they give 
$X^\infty_\pp[u](n=2,3)\simeq 0.55$, which is only $10\%$ above the exact value $1/2$.
To confirm this expectation we compute the constrained FDR
\be
X^\infty_\pp[c]=\frac12+\left(1-\frac\e2\right)\left[
\frac16-\left(\frac{49 + 6 n}{108 (8 + n)}-\frac1{12} \right)\e\right]\,,
\label{Xsc}
\ee 
which gives, in three dimensions,
\be
X^\infty_\pp[c](n=2,3)=0.597\,.
\ee
This estimate is indeed quite close to the unconstrained one, 
suggesting that in 
this case the one-loop expression provides more reliable results than
it does for the longitudinal FDR.
In Fig.~\ref{Xinfs} (right panel) 
a comparison between the various estimates as functions of
the dimensionality $d$ is provided for $2\le d \le 4$. 
The unconstrained ones for $n=2,3$, and $\infty$ (denoted by $n=2 [u]$, $n=3
[u]$, and $n =\infty [u]$, respectively) almost coincide due to the weak
dependence on $n$ of the expression~\reff{Xinfpexpr}. As a consequence,
the corresponding constrained estimates ($n=2 [c]$, $n=3
[c]$, and $n =\infty [c]$, respectively) are not distinguishable on
this scale. For comparison we report (dashed line) 
the exact prediction in the limit $n\rightarrow\infty$ (see
Eq.~\reff{Xinfninf}) and the prediction~\reff{XinfFT} for the
$\tilde\epsilon$-expansion ($\tilde\epsilon = d-2$) close to the lower
critical dimension for $n=3$ and $n=\infty$ (denoted by $n=3
[\tilde\epsilon]$ and $n=\infty [\tilde\epsilon]$, respectively). 
As already pointed out in Sec.~\ref{Secninf} the result $n=\infty
[\tilde\epsilon]$ is indeed tangential to the exact result
$n=\infty, {\rm exact}$ (dashed line, Eq.~\reff{Xinfninf}). The
constrained estimate $n=3 [c]$ provided by Eq.~\reff{Xsc} does not
account completely for the behavior of
$X^\infty_\s(n=\infty)$ close to $d=2$, as predicted by
Eq.~\reff{XinfFT}, but reproduces only its value for $d=2$. 
Reliable interpolation formulas accounting properly for both 
the behaviors close to $d=2$ and $d=4$ require the knowledge of higher
order terms in $\tilde\e$ and $\e$ than those currently available.

\section{Conclusions}
\label{sec-con}

In this paper we have studied the non-equilibrium dynamics of a system
belonging to the $O(n)$ universality class and relaxing at criticality
according to a purely dissipative dynamics from a state with initial 
magnetization $M_0$ (induced via an external magnetic field). 
In the aging regime ($t\gg s\gg \tau_m \sim M_0^{-1/\k}$) we found that 
the  zero-momentum transverse ($\pp$) and longitudinal ($\s$) response 
and {\it connected} correlation functions have the scaling forms
\bea
R_{\q=0}^\s(t,s) &=& A_R^\s\, (t-s)^a(t/s)^{\theta_\s} F_R^\s
(s/t)\; ,\\
C_{\q=0}^\s(t,s) &=& A_C^\s\,s(t-s)^a(t/s)^{\theta'_\s} F_C^\s
(s/t)\; ,\\
R_{\q=0}^\pp(t,s) &=& A_R^\pp\, (t-s)^a(t/s)^{\theta_\pp}
F_R^\pp (s/t)\; ,\\
C_{\q=0}^\pp(t,s) &=& A_C^\pp\,s(t-s)^a(t/s)^{\theta'_\pp}
F_C^\pp (s/t)\; ,
\eea
independently of the actual value of $M_0\neq0$.
We argued that the exponents $\theta_{\s,\pp}$ and $\theta'_{\s,\pp}$ are not 
new non-equilibrium quantities, but they can be written in terms of 
equilibrium exponents as
\bea
\theta_\s &=& \theta'_\s = - \frac{\beta\delta}{\nu z}\,,\\
\theta_\pp&=& \theta'_\pp= - \frac{\beta}{\nu z}\,.
\eea
The two response and correlation functions define two (in principle different)
FDRs $X^\infty_\s$ and $X^\infty_\pp$.

We solved exactly the dynamics of the transverse modes within the Gaussian 
approximation and we derived all the relevant universal ratios and scaling 
functions (the dynamics of the longitudinal modes was already
discussed in 
Ref.~\cite{cgk-06}). In particular we found that the FDRs of longitudinal and
transverse modes are different, making the effective temperature
(defined via $X^\infty$) observable-dependent already within the Gaussian
approximation.
Then we solved exactly the dynamics of the transverse modes in the limit 
$n\to\infty$. The resulting FDR interpolates correctly (as a 
function of the aging time) from the well-known spherical model value 
$X^\infty=1-2/d$ \cite{gl-00c} at $M_0=0$ and the new result 
\be
X^\infty_\pp=\frac{d}{d+2}\,,
\ee
at any $M_0\neq0$. 
This result coincides with the one for local spin observables in the spherical
model \cite{as-05}.

In order to provide 
more accurate predictions in physical dimensions $d=2,3$ for
the physical models with $n=2,3$ we computed the first-order correction in the 
$\e$-expansion to the equation of motion, to the response and correlation 
functions and from them we obtained the FDRs. 
In particular for the longitudinal modes we found
\be
X^\infty_{\s}=\frac{4}{5}-\frac{1895+76 n-162 \pi^2}{1800(8+n)}\e +O(\e^2)\,,
\ee
which differs for $n=\infty$ from the $\e$-expansion of the
corresponding spherical model
result~\cite{as-05}. 
To our knowledge this is the first example of a 
universal quantity taking different values in 
these two models, which are known to be 
equivalent in equilibrium~\cite{stanley-68}.
It is not yet clear to us whether this discrepancy
is actually due to an effectively different 
initial state or to non-Gaussian corrections to the action which have
to be taken into account for the longitudinal mode (see Ref.~\cite{as-05}).
For comparison, it would be interesting to calculate the leading non-Gaussian 
contributions to the $O(\infty)$ model.
For finite $n$ we provided an estimate of $X^\infty_{\s}$ by
constraining it to assume the known value in two dimensions (see
Eq.~(\ref{XinfLc})), resulting in $X^\infty_\s(n=2,3)\simeq 0.71$ in
three dimensions.
For the transverse FDR we obtained
\be
X^\infty_{\pp}=\frac{2}{3}-\frac{49 + 6 n}{108 (8 + n)}\e +O(\e^2)\,.
\ee 
The constrained estimates (see Eq.~\reff{Xsc}) in $d=3$ yield
$X^\infty_\pp(n=2,3)\simeq 0.60$. 

An interesting extension of the work presented here would be to
consider the non-equilibrium evolution at criticality of the $O(n)$ model
starting from an equilibrium magnetized state which has been 
prepared in the low-temperature phase $T<T_c$ in zero
magnetic field. In this case the transverse modes are critical even in the 
initial state, affecting in a non-trivial way the ensuing evolution.

\section*{Acknowledgments}

We thank P. Sollich  for very useful discussions. 
PC acknowledges the financial support from the 
Stichting voor Fundamenteel Onderzoek der Materie (FOM).

\appendix

\section{Determination of the \htit function}
\label{App-h}

In this appendix we determine the function $h(x)$ 
[introduced in Eq.~\reff{eq-defh}] which 
characterizes completely the dynamics of the transverse modes in the 
limit $n\to\infty$.

First of all 
we express the self-consistency condition Eq.~\reff{eq-gap} 
in terms of $h$ only. To this end we note that Eq.~\reff{eq-defh} implies:
\be
\hat r(t) =\L^2 \frac{h'(2\L^2 t)}{h(2\L^2 t)}\,,
\ee
and that, from Eqs.~\reff{eq-R-h} and~\reff{eq-expr-C},
\be
\int^\L\!\!(\rmd q) C_\q^\pp(t,t) = 2 \int_0^t\!\!\rmd t'
\, \frac{h(2\L^2t')}{h(2\L^2t)}\int^\L\!\!(\rmd q) \rme^{-2\q^2(t-t')}
= 2 N_d \L^d \int_0^t\!\!\rmd t'
\, \frac{h(2\L^2t')}{h(2\L^2t)} K_d(2\L^2(t-t'))\,,
\ee
where we have introduced the kernel 
\be
K_d(\tau) \equiv N_d^{-1} \L^{-d} \int^\L\!\!(\rmd q)\, \rme^{-\q^2\tau/\L^2}
\,.
\label{eq-defK}
\ee
Its explicit expression depends on the specific implementation of
the large-momentum cut off. For later convenience we have introduced
the factor $N_d = 2/[(4 \pi)^{d/2} \Gamma(d/2)]$ in the definition of $K_d$.
Then we compute
\be
\int^\L\!\!(\rmd q) C_\q^\pp(\infty,\infty) =  \int^\L\!\!(\rmd q)
\frac{1}{\q^2 + \hat r_\infty} 
= N_d \L^{d-2} I_d(\hat r_\infty/\L^2)\,,
\ee
where $I_d(z)$ is given by 
\be
I_d(z) \equiv N_d^{-1} \L^{-(d-2)} \int^\L\!\!(\rmd q) \frac{1}{\q^2 + \L^2 z}
\label{eq-defI}
\ee
($I_d(0)$ is finite for $d>2$).
Using these results and Eq.~\reff{eq-m-h}, the consistency 
condition~\reff{eq-gap} can be 
expressed in terms of $h$ only 
\be
\L^2\frac{h'(\tau)}{h(\tau)} = \hat r_\infty + \frac{gN_d}{6} \L^{d-2}
\left\{\int_0^\tau\!\!\rmd \tau'
\, \frac{h(\tau')}{h(\tau)} K_d(\tau-\tau') -   I_d(\hat
r_\infty/\L^2) \right\} + \frac{m_0^2}{6} \frac{1}{h(\tau)}\,,
\ee
where $\tau = 2\L^2 t$ and $\tau' = 2\L^2 t'$ are 
dimensionless time variables.
This equation turns into a linear differential equation for $h$:
\be
h'(\tau) = \frac{\hat r_\infty}{\L^2} h(\tau) + u
\left\{\int_0^\tau\!\!\rmd \tau'
\, h(\tau') K_d(\tau-\tau') -   I_d(\hat
r_\infty/\L^2) h(\tau) \right\} + \frac{1}{6}\frac{m_0^2}{\L^2}\,,
\label{eq-gap-h}
\ee
where the dimensionless coupling constant is given by $u \equiv  g N_d
\L^{d-4}/6$ and which can be easily solved by using the Laplace transform 
$\hat g(s) = \La[g(\tau)] \equiv \int_0^\infty \!\! \rmd\tau g(\tau) \rme^{- s\tau}$, 
defined in the half-plane ${\rm Re} (s) > \gamma$, where $\gamma$ depends
on the analytical properties of $g$.
Once $\hat g(s)$ is known, the original function is obtained via the
inverse transform
$g(\tau) = (2\pi i)^{-1} \int_{\gamma' - i\infty}^{\gamma' + i\infty} 
\rmd s \; \hat g(s) \rme^{s\tau}$ for $\gamma' > \gamma$.
Transforming
Eq.~\reff{eq-gap-h} one gets $\La[h'(\tau)] = - h(0) + s \hat h(s)$
where, from the definition~\reff{eq-defh}, $h(0) = 1/2$, and
\be
\int_0^\infty \!\!\rmd\tau\, \rme^{- s\tau} \int_0^\tau\!\!\rmd \tau'
\, h(\tau') K_d(\tau-\tau') = \int_0^\infty \!\!\rmd\tau' \rme^{-
s\tau'} h(\tau')
\int_{\tau'}^\infty\!\!\rmd \tau  \rme^{- s(\tau - \tau')}
K_d(\tau-\tau') 
= \hat h(s) \hat K_d(s)\,,
\ee
where (see Eq.~\reff{eq-defK})
\be
\hat K_d(s) \equiv \int_0^\infty\!\!\rmd \tau\,   \rme^{- s\tau}
K_d(\tau)
=  \Lambda^{-d} \int_0^\infty\!\!\rmd \tau\,  \rme^{- s\tau} 
\int^\L\!\!(\rmd q)\rme^{-\q^2\tau/\L^2}
=  \Lambda^{-d} \int^\L\!\!(\rmd q) \frac{1}{\q^2/\L^2 + s} 
= I_d(s)\,.
\ee
Accordingly, Eq.~\reff{eq-gap-h} becomes
\be
-\frac{1}{2} + s \hat h(s) = \frac{\hat r_\infty}{\L^2} \hat h(s) +
u \left[I_d(s) - I_d(\hat r_\infty/\L^2)\right] \hat h(s) +
\frac{1}{6} \frac{m_0^2}{\L^2} \frac{1}{s}\,,
\ee
whose solution is
\be
\hat h(s) =\frac{\displaystyle \frac{1}{2} + \frac{1}{6}\frac{m_0^2}{\L^2}
\frac{1}{s}}{\displaystyle s - \frac{\hat r_\infty}{\L^2} - 
u [I_d(s) - I_d(\hat r_\infty/\L^2)]}\,.
\label{eq-solh}
\ee
First we note that, as expected, $\hat h(s)$ has a pole for $s
= s_\infty \equiv \hat r_\infty/\L^2$ independently of the value of $u$.
For $u=0$, $\hat h(s)$ is analytic in the complex plane
except for the poles at $s=s_\infty$ and $s=0$. Accordingly, the
corresponding inverse
Laplace transform is
\be
h(\tau) = \frac{1}{2}\rme^{s_\infty\tau}
+\frac{1}{6}\frac{m_0^2}{\L^2}\frac{1}{s_\infty} \left(\rme^{s_\infty
\tau} - 1\right)\,,
\ee
i.e.,
\be
h(2\Lambda^2t) =  \frac{1}{2}\rme^{2\hat r_\infty t}
+\frac{1}{6}\frac{m_0^2}{\hat r_\infty} 
\left(\rme^{2\hat r_\infty t} - 1\right)\,,
\ee
which is actually independent of $\L$, as expected. From this
expression one recovers all the Gaussian results of
Sec.~\ref{sec-Gaux} in the limit $\hat r_\infty \to 0$ (corresponding
to the critical point).
\begin{figure}[t]
\begin{center}
\epsfig{width=7truecm,file=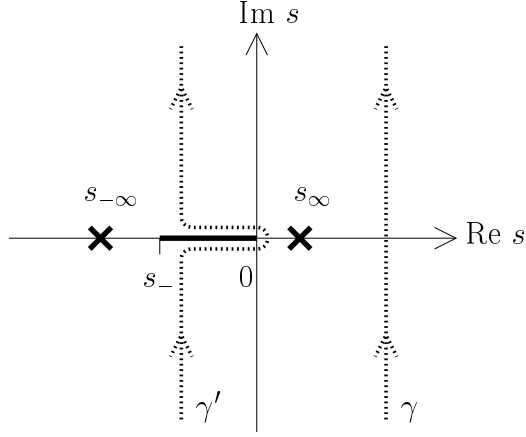}
\end{center}
\caption{Analytic structure of $\hat h(s)$ and contour of integration for 
the computation of its inverse Laplace transform. 
Poles are indicated by a cross, whereas
the branch cut by the thick solid line. $\gamma$ is the original
contour of integration to compute the inverse transform of $\hat
h(s)$, whereas $\gamma'$ is its convenient deformation in the case
$s_\infty =0$ (see Appendix~\ref{app-poles}).}
\label{contour}
\end{figure}

Because of $I_d(s)$, $\hat h(s)$ has a branch cut on the
real axis for $s_-<\mbox{Re}(s)<0$ and $u>0$, in addition to the poles for
$s=s_\infty$ and $s=0$. If $s_-$ is finite then 
an isolated pole $s_{-\infty}$ is also present on the real
axis with $\mbox{Re}(s_{-\infty}) < s_-$ (see Appendix~\ref{app-poles}).
In the scaling limit we are interested in, $\L^2 t \gg 1$ ($\tau \gg
1$), the relevant contribution
in the inverse Laplace transform comes from the singularities with the
largest real part.
If $\hat r_\infty > 0$, this is given by
$s=s_\infty$, leading, as in the Gaussian case, to an expected exponential
decay of $h^{-1}(\tau)$.
For $\hat r_\infty = 0$, instead,
\be
\hat h(s) =\frac{\displaystyle \frac{1}{2} + \frac{1}{6}\frac{m_0^2}{\L^2}
\frac{1}{s}}{\displaystyle s -
u [I_d(s) - I_d(0)]}\,,
\label{eq-solh-crit}
\ee
and the leading singularity comes from the branch cut.
Using the asymptotic expansion for $I_d(s)$ derived in Appendix~\ref{app-exp}
we can write the denominator of Eq.~(\ref{eq-solh-crit}) as 
\be
s - u [I_d(s) - I_d(0)] =
 a_0 s^{d/2-1} + b_0 s + \sum_{k=1}^\infty \left[a_k s^{d/2-1 + k} +
b_k s^{1+k}\right]\,,
\label{exp-II}
\ee
where
\be
a_0 = u\kappa_d \quad \mbox{and}\quad b_0 = 1 -u\, \alpha_d\,,
\label{eq-coeff}
\ee
and $\kappa_d=\pi/[2 \sin(\pi (d-2)/{2})]$ is a universal coefficient, \ie\ 
independent of the actual
regularization procedure, whereas $\alpha_d$ does depend on it (with
the universal feature $\alpha_{4-\e} \sim 1/\e$ for $\e \rightarrow 0$). 
Accordingly, for $s\to 0$ and $2<d<4$ (so that $\e = 4-d >0$) one has
\be
\frac{1}{s - u [I_d(s) - I_d(0)]} = 
a_0^{-1} s^{-1 +\e/2}  \left[1 + \sum_{m,n=0}^\infty 
c_{m,n} s^{m + n\e/2} \right]\,,
\ee
where $c_{0,0} = 0$,  
$c_{0,n} = (- b_0/a_0)^n$ and all the other coefficients can be
computed in terms of $a_k$ and $b_k$. [Note that for $d>4$ the leading
behavior for $s\to 0$ is, apart from a coefficient, the same as in the
case $u=0$, leading to the expected Gaussian-like behavior.]
The inverse Laplace transform of $\hat h(s)$ for $\tau \gg 1$ is
dominated by the behavior of $\hat h(s)$ for $s\to 0$, and it can be computed
following a suitable contour of integration in the complex plane
(indicated as $\gamma'$ in Fig.~\ref{contour}) on
the two sides of the branch cut. Recalling that 
$\La^{-1}[1/s^\alpha] = \tau^{\alpha-1}/\Gamma(\alpha)$,
one concludes that 
\bea
h(\tau \gg 1) &=& \frac{a_0^{-1}}{2}\left[
\frac{\tau^{-\e/2}}{\Gamma(1-\e/2)} +
\sum_{n=1}^{\lfloor\frac{2}{\e}\rfloor - 1}\frac{c_{0,n}}{\Gamma(1
-(n+1)/2\e)} \tau^{-(n+1)/2\e}\right] \\
&+& \frac{a_0^{-1}}{6}\frac{m_0^2}{\L^2} \left[
\frac{\tau^{1-\e/2}}{\Gamma(2-\e/2)} + 
\sum_{n=1}^{\lfloor\frac{4}{\e}\rfloor - 1}\frac{c_{0,n} 
\tau^{1-(n+1)/2\e}}{\Gamma(2-(n+1)/2\e)} +
\sum_{n=1}^{\lfloor\frac{2}{\e}\rfloor - 1}\frac{c_{1,n} \tau^{-(n+1)/2\e}}{
\Gamma(1 -(n+1)/2\e)} \right]\nonumber\\
&+& O(\tau^{-1})\,.\nonumber
\eea
The confluent corrections $\sim \t^{-n\e/2}$ ($n>2$) to the leading scaling 
behavior vanish identically if $c_{0,n} = 0$, i.e., for $b_0 = 0$. 
According to Eq.~\reff{eq-coeff}
this occurs if $u=u^*= \alpha_d^{-1}$, which is indeed 
the fixed-point value for 
$u$ (note that if dimensional regularization is employed, this leads to 
the well-known result $u^*=\e$).  

Accordingly, the scaling limit (\ie\ $\L\to\infty$ with  $m_0^2 t$
fixed) for $2<d<4$ is characterized by the fixed-point expression
\be
h(2\L^2 t \gg 1) =
\frac{1}{u^*} \frac{\kappa_d^{-1}}{2\Gamma((d-2)/2)} (2 \L^2 t)^{-\e/2}
\left[ 1 + \frac{4}{3(d-2)}m_0^2 t + O((2\L^2 t)^{-\e/2})\right]\,,
\ee
which is the result reported in Eq.~(\ref{h-sol}).

\subsection{Poles of $\hat h(s)$}
\label{app-poles}

Here we investigate the properties of analyticity for $s \in \mathbb C$ of
\be
\hat h(s) = 
\frac{\displaystyle \frac{1}{2} + \frac{1}{6}\frac{m_0^2}{\L^2}
\frac{1}{s}}{\displaystyle s - \rho -
u [I_d(s) - I_d(\rho)]}\,,
\ee
where $\rho \equiv \hat r_\infty/\L^2> 0$. Clearly,
the numerator has a pole for $s=0$ whereas the denominator vanishes for 
$s = \rho$, independently of $u>0$. 
$I_d(s)$ [see Eq.~\reff{eq-defI}] has a branch
cut on the real axis for 
$s_- \equiv -\q_{\rm max}^2/\L^2<\mbox{Re}\; s < 0$, where $\q_{\rm max}$
is the maximum momentum in the integration domain ($|\q_{\rm max}|\sim \L$). 
Depending on the actual definition of the momentum
integral $\int^\L(\rmd q)$, $s_-$ might be not finite.
Possible additional poles are given by
the solution of
\be
 s - \rho -u [I_d(s) - I_d(\rho)] = 0\,.
\ee
Note that, as a consequence of Eq.~\reff{eq-defI},
\be
I_d(s) - I_d(\rho) = - (s - \rho) N_d^{-1} \L^{4-d}\int^\L\!(\rmd q) 
\frac{1}{(\q^2 + \L^2 s)(\q^2 + \L^2 \rho)}\,,
\ee
and therefore additional poles are given by the solutions of
\be
J_d(s) \equiv N_d^{-1} \L^{4-d} \int^\L\!(\rmd q) 
\frac{1}{(\q^2 + \L^2 s)(\q^2 + \L^2 \rho)} = -\frac{1}{u}\,.
\label{eq-pole}
\ee
It is easy to see that the imaginary part of the integral vanishes
only on the real axis and therefore possible poles have to be
located there and we can restrict our search to $s \in (-\infty,s_-)$
given that $J_d(s\in {\mathbb R}^+) > 0$. Note that
$J_d(s<s_-) \le 0$, $J_d(s\rightarrow -\infty) = 0^-$, and that
$J'_d(s) < 0$ (as it is clear by differentiating the integral). In addition,
$J_d(s\rightarrow s_-)$ diverges logarithmically and therefore 
Eq.~\reff{eq-pole} admits a unique solution $s_{-\infty}< s_-$ for
$u\neq 0$.
In the case of a regularization with a sharp cut-off $\L$ one finds
$s_-=-1$ and it
is easy to determine the asymptotic behavior of $J_d(s)$ for $s
\rightarrow -1^-$: $J_d(s\rightarrow -1^-) = -
(1+\rho)^{-1}\ln (-1-s) + \mbox{finite}$ and therefore
Eq.~\reff{eq-pole} admits a unique solution $s_{-\infty}<-1$ for
$u\neq 0$. In particular, for $u$ small one has $s_{-\infty} \simeq -1
- \rme^{-2(1+\rho)/u}$.

\subsection{Expansion of $I_d(s)$}
\label{app-exp}

\newcommand{\cf}{\varphi}

Let us consider a possible definition of the regularized integral, via
a cut-off function $\cf(x)$ (defined for $x\in {\mathbb R}^+$), i.e.,
\be
\int^\L\!\!(\rmd q) \mapsto \int_{{\mathbb R}^d}\!(\rmd q)\, \cf(\q^2/\L^2)\,,
\ee
where we assume that $\cf$ is a smooth function which 
does not change the small-momentum behavior of the integrand, i.e.,
$\cf(0) = 1$ and decays exponentially fast for $x\rightarrow\infty$.
For later convenience
we also assume that $\cf(x)$ can be expanded around $x=0$ (or, at
least that $\cf'(0)$ is finite). These properties are also encoded in
the analytic properties of the Mellin transform
$\bar\cf(s)$ of $\cf(x)$, defined as
\be
\bar \cf(z) \equiv \int_0^\infty\!\!\rmd x\, x^{z-1} \cf(x)\,,
\ee
which converges in the complex plane for $\mbox{Re}(z) > 0$
(in case of an algebraic decay of $\cf(x)$ for $x\rightarrow\infty$ this
domain becomes a strip). Its analytic extension to $z\in \mathbb C$
is characterized by poles on the real axis 
for the non-positive integers $n = 0,-1,-2,\ldots$ with residues 
$\cf^{(n)}(0)/n!$, due to the small-$x$ behavior of $\cf(x)$.
The original function $\cf(x)$ is recovered by computing
\be
\cf(x) = \frac{1}{2\pi i} \int_{c-i\infty}^{c+i\infty} \!\!\rmd z\, \bar\cf(z) x^{-z} \,,
\ee  
with $c>0$.

Using this representation of $\cf$ in the definition of $I_d$ [see
Eq.~\reff{eq-defI}] we find 
\bea
I_d(s) 
=\frac{1}{2\pi i } \int_{c-i\infty}^{c+i\infty} \!\!\rmd z\,
\bar\cf(z) \,s^{d/2-1-z} \frac{\pi}{2 \sin \pi(d/2-z)}\,,
\eea
where $d/2-1<c<d/2$.
This expression allows us to determine the asymptotic behavior
for $s\rightarrow 0$, by shifting the integration contour towards
negative values of $\mbox{Re}\; z$:
\be
I_d(s) = \frac{1}{2}\bar\cf(d/2-1) + \frac{\pi}{2 \sin \left(\pi d/2\right)}
\cf(0) s^{d/2-1} -\frac{1}{2} \bar\cf(d/2-2) s +
\sum_{k=1}^\infty \left[ A_k s^{d/2-1+k}+ B_k s^{1+k}\right]\,.
\label{app-fexp}
\ee
Note that the first term, $I_d(0)$ ($d>2$), depends on the specific choice of
the cut-off function $\cf(x)$, whereas the second term is actually 
universal provided that $\cf(0)=1$. Higher-order terms do depend on
$\cf(x)$.
Let us consider the linear term is $s$. 
The analytic extension of $\bar \cf(z)$ to the region $-1<\mbox{Re}(z)<0$ 
is readily computed as
\be
\bar \cf(z) = \int_0^\infty\!\!\rmd x\, x^{z-1}[\cf(x) - \cf(0)]\,,
\ee 
and therefore 
\be
\alpha_d \equiv - \frac{1}{2} \bar \cf(d/2-2) = -\frac{1}{2} \int_0^\infty\!
\rmd x\, x^{d/2-3}[\cf(x) - 1]
\ee
behaves as
\be
\alpha_{4-\e} = \frac{1}{\e} + \mbox{finite} \quad \mbox{for}\quad
\e\rightarrow 0\,.
\ee
Accordingly, Eq.~\reff{app-fexp} can be cast in the form
\be
I_d(s) = I_d(0) - \kappa_d \, s^{d/2-1} + \alpha_d\, s + \sum_{k=1}^\infty \left[ A_k s^{d/2-1+k}+ B_k s^{1+k}\right]\,,
\ee
where $\kappa_d = \pi/[2 \sin (\pi (d-2)/2)]$, which leads 
to Eq.~\reff{exp-II}. 
The same conclusion can be drawn by adopting different regularization
schemes (see, \eg\  Appendix B of Ref.~\cite{mz-03}).

\section{One-loop Feynman diagrams}
\label{app-conti}

As usual in the computations of Feynman diagrams, it is convenient to
calculate first the corresponding one-particle irreducible (1PI) parts.
For the response functions
they are given by the two tadpoles $I_\s$, $I_\pp$ (see Eqs.~\reff{Is} and \reff{Ip})
and by the four ``bubbles'' which can be written as (hereafter the
response and correlation functions appearing in the integrals are 
the Gaussian ones)
\bea
B_{LL}(t,s)&=& \int (\rmd q) R_\q^\s(t,s)C_\q^\s(t,s)=
\frac{2 t^{-3}}{(8\pi)^{d/2}}\int_0^s \rmd x\, x^3 (t-x)^{-d/2}\equiv
\frac{2}{(8\pi)^{d/2}} H_L(t,s)\,,\\
B_{TL}(t,s)&=& \int (\rmd q) R_\q^\pp(t,s)C_\q^\s(t,s)=\frac{t}{s} B_{LL}(t,s)=
\frac{2}{(8\pi)^{d/2}} \frac{t}{s}H_L(t,s)\,,\label{BTL}\\
B_{TT}(t,s)&=& \int (\rmd q) R^\pp_\q(t,s)C^\pp_\q(t,s)=
\frac{2 t^{-1}}{(8\pi)^{d/2}}\int_0^s \rmd x\, x (t-x)^{-d/2}\equiv
\frac{2}{(8\pi)^{d/2}} H_T(t,s)\,,\label{BTT}\\
B_{LT}(t,s)&=& \int (\rmd q) R^\s_\q(t,s)C^\pp_\q(t,s)=\frac{s}{t} B_{TT}(t,s)=
\frac{2}{(8\pi)^{d/2}} \frac{s}{t} H_T(t,s)\label{BLT}\,,
\eea
where
\bea
H_L(t,s)&=&t^{-3} \int_0^s \rmd x\, x^3 (t-x)^{-d/2}
\nonumber\\&=&
A_L t^{-1+\e/2}-A_L t^{-1}(t-s)^{\e/2}+B_L t^{-2}s (t-s)^{\e/2}
+C_L t^{-3}s^2 (t-s)^{\e/2} \nonumber\\&&
+D_L t^{-3}s^3 (t-s)^{-1+\e/2}\,,\label{paraL}\\
H_T(t,s)&=&t^{-1} \int_0^s \rmd x\, x (t-x)^{-d/2}=
\nonumber\\&=&
A_T t^{-1+\e/2}-A_T t^{-1} (t-s)^{\e/2}+B_Tt^{-1} s (t-s)^{-1+\e/2}\,,
\label{para}
\eea
with
\bea
A_L&=&\frac{6}{(2+\e/2)(1+\e/2)\e/2(-1+\e/2)}\,,\\
B_L&=&-\frac{6}{(2+\e/2)(1+\e/2)(-1+\e/2)}\,,\\
C_L&=&-\frac{3}{(2+\e/2)(-1+\e/2)}\,,\\
D_L&=&-\frac{1}{(-1+\e/2)}\,,\\
A_T&=&\frac{1}{\e/2(-1+\e/2)}\,,\label{c-AT}\\
B_T&=&-\frac{1}{(-1+\e/2)}\,.
\eea
This parameterization of the 1PI parts of the integrals is particularly
convenient, as explained in Ref.~\cite{cgk-06}.

In what follows we shall make extensive use of the 
results of Ref.~\cite{cgk-06}. 
However, we alert the reader to the fact that the factor 
$r_d$ appearing in Ref.~\cite{cgk-06} has to be identified, in the
current notation, with  
$r_\s$ and {\it not} with $r_d \equiv r_\s + (n-1) r_\pp/3$ 
(unless differently stated).

\subsection{Longitudinal response}
\label{appRL}

The integral $I_1$ and $I_2$ have been 
computed in Ref.~\cite{cgk-06}, Eq.~(A.9) and~(A.11), respectively. 
$I_3$ is easily related to $I_1$:
\be
I_3(t,s)=\int_s^t \rmd t' R^\s(t,t')  I_\pp(t') R^\s(t',s)
= \frac{r_\pp}{r_\s} I_1(t,s) \label{I3}\,,
\ee
whereas $I_4$ is given by
\bea
I_4(t,s)&=&\int_s^t \rmd t_1 \int_s^{t_1} \rmd s_1 R^\s(t,t_1)
\sqrt{2} m(t_1) B_{TT}(t_1,s_1)\sqrt{2} m(s_1)  R^\s(s_1,s)\nonumber\\
&=& \frac{6}{(8\pi)^{d/2}}  \left(\frac{s}{t}\right)^{3/2}
\int_s^t \rmd t_1\, t_1 \int_s^{t_1} \rmd s_1\, s_1^{-2} H_{T}(t_1,s_1)\,.
\eea
According to the decomposition~\reff{para} of $H_T$,
this expression can be written as $I_4=\sum_{i=1}^3 (a_i)$, where $(a_i)$
are easily calculable:
\bea
\frac{(a_1)}{A_T}&=&\int_s^t \rmd t_1\int_s^{t_1}\rmd s_1\,
s_1^{-2}t_1^{\e/2}
= \frac{t^{1+\e/2}-s^{1+\e/2}}{s(1+\e/2)}-\frac2\e(t^{\e/2}-s^{\e/2})
\nonumber\\
-\frac{(a_2)}{A_T}&=&\int_s^t \rmd t_1\int_s^{t_1}\rmd s_1\, s_1^{-2}
(t_1-s_1)^{\e/2} \nonumber\\
&=& \frac1{1+\e/2}\left[\frac{t}{s}-1-\ln \frac{t}{s}+\frac\e2
\left(\left(\frac{t}{s}-1\right) \ln (t-s)-(1+\ln t)\ln \frac{t}{s}
+\frac{\pi^2}{6} -\Li_2(s/t)\right)\right] + O(\e)\nonumber\,,\\
\frac{(a_3)}{B_T}&=&\int_s^t \rmd t_1\int_s^{t_1}\rmd s_1\,
s_1^{-1}(t_1-s_1)^{-1+\e/2}
= \frac2\e\ln \frac{t}{s}+\ln t \ln
\frac{t}{s}+\Li_2(s/t)-\frac{\pi^2}6 + O(\e)\,.
\nonumber
\eea
and whose sum is ($x\equiv s/t$)
\be
I_4= \frac{6s^{\e/2}}{(8\pi)^{d/2}} x^{3/2}
\left[-\frac{2}{\e}\ln x+\frac{\ln^2 x}2+\left(\frac1x-1\right)\ln(1-x)+O(\e)
\right]\,.
\ee

\subsection{Longitudinal correlation}
\label{appCL}

The integrals $I_6$ and $I_7$ have been calculated in
Ref.~\cite{cgk-06}, Eq~(B.25) [see also Eq.~(B.26)] and Eq.~(B.27),
respectively, whereas $I_{10}$ is trivially related to $I_7$: 
\be
I_{10}=\frac{r_\pp}{r_\s} I_7\,.
\ee
$I_9$ can be written as
\bea
I_9(t,s) &=& \int_0^t\!\rmd t_2 \int_{t_2}^t\!\rmd t_1\, R^\s_{\q
=0}(t,t_1) \sqrt{2} m(t_1) B_{TT}(t_1,t_2) \sqrt{2} m(t_2)  C^\s_{\q
=0}(t_2,s) + \nonumber\\
&&\quad +  \int_0^s\!\rmd t_2 \int_{t_2}^s\!\rmd t_1\, C^\s_{\q
=0}(t,t_1) \sqrt{2} m(t_1) B_{TT}(t_2,t_1) \sqrt{2} m(t_2)  R^\s_{\q
=0}(s,t_2)
\eea
which,  using Eq.~\reff{BTT}, becomes
\bea
I_9(t,s) &=& \frac{3}{(8 \pi)^{d/2}} (s\,t)^{-3/2} 
\left[ \int_0^s\rmd t_2\, t_2^2 \int_{t_2}^t\!\rmd t_1\, t_1
H_T(t_1,t_2) \right. \nonumber \\
&& \quad \left. + s^4
\int_s^t\!\rmd t_2 \, t_2^{-2} \int_{t_2}^t \!\rmd t_1\, t_1
H_T(t_1,t_2) +
\int_0^s\!\rmd t_1\, t_1^2 \int_{t_1}^s\!\rmd t_2\, t_2\,
H_T(t_2,t_1)\right].
\eea
We denote the three terms in square brackets by $(A)$, $(B)$, and
$(C)$, i.e., $[\ldots] = [(A)+(C)+(C)]$. In
turn each of them can be written as the sum of the three
contributions coming from the decomposition of $H_T$ according to
Eq.~\reff{para}, leading to the following expression for $I_9$:
\be
I_9(t,s)  =
\frac{3(st)^{-3/2}}{(8\pi)^{d/2}} \sum_{b\in{A,B,C}} \sum_{i=1}^3 (a)_i\,.
\ee
The terms with $i=1$ and $2$ are the same as the
corresponding ones calculated in 
Ref.~\cite{cgk-06}, Eqs.~(B.10),
(B.11), (B.15), (B.16), (B.20), and (B.21), with the proviso that
$A_d$ therein has to be replaced by $A_T$ here (see Eq.~\reff{c-AT}). 
We still need 
\bea
\frac{(A)_3}{B_T}&=&\int_0^s\rmd t_2\,t_2^3\int_{t_2}^t \rmd t_1\, (t_1-t_2)^{-1+\e/2}\nonumber\\&=&
\frac{s^4}{2\e} -\frac{s^4}{16}-\frac{s^3t}{12}-\frac{s^2t^2}8-\frac{s t^3}4-
\frac14(t^4-s^4)\ln(t-s)+\frac{t^4}4\log t+O(\e)\,,\\
\frac{(B)_3}{B_T s^4}&=&\int_s^t \rmd t_2\,t_2^{-1}\int_{t_2}^t \rmd t_1\, (t_1-t_2)^{-1+\e/2}=
\frac2\e\ln\frac{t}s+\ln t \ln\frac{t}s+\Li_2(s/t)-\frac{\pi^2}6+O(\e)\,,\\
\frac{(C)_3}{B_T}&=&\int_0^s\rmd t_1\,t_1^3\int_{t_1}^s \rmd t_2\, (t_2-t_1)^{-1+\e/2}=
s^4\left(\frac1{2\e}-\frac{25}{48}+\frac14\ln s\right)+O(\e)\,.
\eea
Summing these nine terms we have
\be
I_9(t,s)  =
\frac{3s^{4+\e/2}(st)^{-3/2}}{(8\pi)^{d/2}}
\left[
\frac{1-2\ln x}{\e}+ \frac{\ln^2 x}{2}-\frac{107}{72}+{\cal C}(x)
\right]\,,
\ee
where
\be
{\cal C}(x)= \frac43 \frac{\ln (1-x)}{x} - \frac{1}{3 x^4} \left[\ln (1-x)
+ x + \frac{x^2}{2} + \frac{x^3}{3}\right] + \frac{5}{4} - \ln (1-x)
\label{Cx}
\ee
[${\cal C}(0)=0$].

$I_5$ has been calculated in Ref.~\cite{cgk-06}, Eqs.~(B.5) and (B.6).
In order compute $I_8$ we first need its 1PI part  [we
define $t_< = \min\{t,t'\}$]
\bea
B_{CC}^{TT}(t,t')&=&\int (\rmd q) [C^\pp_\q(t,t')]^2= \int (\rmd q)
4(tt')^{-1} \rme^{-2 q^2(t+t')} \int_0^{t_<} \rmd t_1 \, t_1 \rme^{2 q^2t_1}
\int_0^{t_<} \rmd t_2 \, t_2 \rme^{2 q^2t_2}=
\nonumber\\&=&
\frac{4(tt')^{-1}}{(8\pi)^{d/2}}\int_0^{t_<} \rmd t_1\int_0^{t_<} \rmd t_2
\frac{ t_1 t_2}{(t+t'-t_1-t_2)^{d/2}}\,.
\eea
so that
\bea
I_8&=&\int_0^s \rmd\t_1\int_0^t \rmd\t_2 \, R^\s(s,\t_1)\sqrt2 m(\t_1)
B_{CC}^{TT}(\t_1,\t_2)\sqrt{2} m(\t_2) R^\s(t,\t_2)\nonumber\\&=&
3(st)^{-3/2}\left[2\int_0^s \rmd\t_1\int_0^{\t_1} \rmd\t_2\, \t_1\t_2 B_{CC}^{TT}(\t_1,\t_2) +
\int_0^s \rmd\t_1\int_s^t  \rmd\t_2 \, \t_1\t_2 B_{CC}^{TT}(\t_1,\t_2)\right]
\nonumber\\&=&
3(st)^{-3/2}[2 B+A]\,.
\eea
Both $A$ and $B$ do not diverge for
$d\to4$. Therefore no dimensional regularization is required
and one can calculate them directly in $d=4$:
\bea
A&=&4 (8\pi)^{-d/2}\int_0^s \rmd\t_1\int_s^t \rmd \t_2
\int_0^{\t_1} \rmd t_1\int_0^{\t_1} \rmd t_2
\frac{t_1 t_2}{(\t_1+\t_2-t_1-t_2)^2} + O(\e)
\nonumber\\&=& 4(8\pi)^{-d/2}s^4
\left[{\cal D}(s/t)+\frac23\ln2 -\frac38 \right]
+ O(\e)\,,\\
B&=&4(8\pi)^{-d/2}\int_0^s \rmd\t_1\int_0^{\t_1} \rmd\t_2
\int_0^{\t_2} \rmd t_1\int_0^{\t_2} \rmd t_2
\frac{t_1 t_2 }{(\t_1+\t_2-t_1-t_2)^2} + O(\e)
\nonumber\\&=&
4(8\pi)^{-d/2}s^4 \left(\frac7{24}-\frac13 \ln2\right) + O(\e)\; ,
\eea
where
\bea
{\cal D}(x)&=&
\frac18+\frac1{4x^3}+\frac1{12 x^2}-\frac1{12x}+
\ln(1-x)\left(\frac1{24}+\frac5{24x^4}-\frac1{6x^3}-\frac1{4x^2}+\frac1{6x}\right)\nonumber\\&+&
\ln(1+x)\left(-\frac1{24}-\frac1{24x^4}-\frac1{6x^3}-\frac1{4x^2}-\frac1{6x}\right)
\label{Dx}
\eea
[${\cal D}(0)=0$].
Accordingly,
\be
I_8=\frac{12}{(8\pi)^{d/2}}s \left(\frac{s}t\right)^{3/2}
\left[\frac{5}{24}+{\cal D}(x)\right]\,.
\ee
Summing all these contributions according to Eq.~\reff{CsumS} one finds
that the function $f_C^\s$ introduced in Eq.~\reff{CresultS} is given by
\be
f_C^\s(x)=\frac{3}{2}{\cal A}(x)+\frac34 {\cal B}(x)+\frac{n-1}{12}{\cal C}(x)
+\frac{n-1}{6}{\cal D}(x) \,,\label{fcs}
\ee
where ${\cal A}(x)$, ${\cal B}(x)$ are given by Eqs.~(B.5) and (B.26),
of Ref.~\cite{cgk-06}, whereas  
${\cal C}(x)$ and ${\cal D}(x)$ are given 
by Eqs.~\reff{Cx} and \reff{Dx} of this appendix.

\subsection{Transverse response}
\label{appRT}

The integrals for $I_{11}$ and $I_{12}$ are (see Eq.~(A.9) in
Ref.~\cite{cgk-06}):
\bea
I_{11}(t,s)&=&\int_s^t \rmd t' R^\pp(t,t')  I_\s(t') R^\pp(t',s)
=\frac{t}s I_1(t,s)\,,\\
I_{12}(t,s)&=&\int_s^t \rmd t' R^\pp(t,t')  I_\pp(t')R^\pp(t',s)
=\frac{r_\pp}{r_\s} I_{11}(t,s)\,.
\eea
$I_{13}$ is, instead, given by
\bea
I_{13}(t,s)&=&\int_s^t \rmd t_1 \int_s^{t_1} \rmd s_1 R^\pp(t,t_1)
\sqrt{2} m(t_1) B_{TL}(t_1,s_1)\sqrt{2} m(s_1)  R^\pp(s_1,s)\nonumber\\
&=& \frac{6}{(8\pi)^{d/2}}  \left(\frac{s}{t}\right)^{1/2}
\int_s^t \rmd t_1\, t_1 \int_s^{t_1} \rmd s_1\, s_1^{-2} H_{L}(t_1,s_1)\,,
\eea
which is just $I_2(t,s) t/s$, where $I_2$ has been calculated in
Eq.~(A.11) of Ref.~\cite{cgk-06}. For $I_{14}$ we have
\bea
I_{14}(t,s)&=&\int_s^t \rmd t_1 \int_s^{t_1} \rmd s_1 R^\pp(t,t_1)
\sqrt{2} m(t_1) B_{LT}(t_1,s_1)\sqrt{2} m(s_1)  R^\pp(s_1,s)\nonumber\\
&=& \frac{6}{(8\pi)^{d/2}}  \left(\frac{s}{t}\right)^{1/2}
\int_s^t \rmd t_1\, t_1^{-1} \int_s^{t_1} \rmd s_1\, H_{T}(t_1,s_1)\,,
\eea
which can be split in three parts, according to the
parameterization~\reff{para} of $H_T$. The final result
is
\bea
I_{14}(t,s)&=&\frac{6s^{\e/2}}{(8\pi)^{d/2}}  \left(\frac{s}{t}\right)^{1/2}
\left[-\frac2\e\ln x+\ln x+\frac{\ln^2x}2-\frac{\pi^2}3
\nonumber\right.\\&+&\left.
3(1-x)-
(1-x)\ln(1-x)+2\Li_2(x)+O(\e)\right]\,.
\eea

\subsection{Transverse correlation}
\label{appCT}

As usual, the tadpole integrals $I_{15}$ and $I_{16}$ are easy
\bea
I_{16}(t,s)&=&\int_0^s\rmd t'\, R^\pp(t',s)I_\s(t')C^\pp(t,t')+
\int_0^t\rmd t'\, R^\pp(t',t)I_\s(t') C^\pp(t',s)=\nonumber\\
&=&2N_d r_\s s \left(\frac{s}{t}\right)^{1/2}s^{\e/2}\left\{\frac{2}{2+\e/2}
+\frac{2}{\e}\left[\left(\frac{t}{s}\right)^{\e/2}-1\right]\right\}\,,\\
I_{15}(t,s)&=&\frac{r_\pp}{r_\s} I_{16}(t,s)\,.
\eea
There are two diagrams involving $B_{LT}(t_1,s_1)$:
\bea
I_{17}(t,s)&=&
\int_0^t \rmd t_2 \int_{t_2}^t \rmd t_1\, R^\pp(t,t_1) \sqrt{2} m(t_1) B_{LT}(t_1,t_2)
\sqrt{2} m(t_2) C^\pp(t_2,s)\nonumber\\&+&
\int_0^s \rmd t_1 \int_{t_1}^s \rmd t_2\, C^\pp(t,t_1) \sqrt{2} m(t_1) B_{LT}(t_2,t_1)
\sqrt{2} m(t_2) R^\pp(s,t_2)\,,
\eea
which can be written, via Eq.~\reff{BLT}, as
\bea
I_{17}(t,s)&=\displaystyle
\frac{6(st)^{-1/2}}{(8\pi)^{d/2}}  &\left[
\int_0^s \rmd t_2 \, t_2^2\int_{t_2}^t \rmd \, t_1 t_1^{-1} H_T(t_1,t_2)+
s^2 \int_s^t \rmd t_2 \int_{t_2}^t \rmd t_1 \, t_1^{-1} H_T(t_1,t_2)
\right.\nonumber\\&&\left.+
\int_0^s \rmd t_1 \, t_1^2\int_{t_1}^s \rmd t_2\, t_2^{-1} H_T(t_2,t_1)\right]\,.
\eea
In turn, we decompose each of the three terms in square brackets in
three parts, according to the parameterization~\reff{para} of $H_T$. 
We do not report the details of the calculation, but only the final result
\be
I_{17}(t,s)=\frac{6 s^{1+\e/2}}{(8\pi)^{d/2}}\left(\frac{s}t\right)^{1/2}
\left[
\frac2\e-\frac2\e \ln x+\frac{\ln^2 x}2+\ln x+\frac{23}{18}-\frac{\pi^2}3
+{\mathcal G} (x)+O(\e)\right]\,,
\ee
where
\be
{\cal G}(x)=-\frac13-\frac{20}9 x-\frac23\frac{\ln(1-x)+x}{x^2}
+\frac23 x \log(1-x)+2\Li_2(x) \label{Gx}
\ee
[${\cal G}(0) = 0$].

The expression for $I_{18}$ is
\bea
I_{18}(t,s)&=&
\int_0^t \rmd t_2 \int_{t_2}^t \rmd t_1\, R^\pp(t,t_1) \sqrt{2} m(t_1) B_{TL}(t_1,t_2)
\sqrt{2} m(t_2) C^\pp(t_2,s)\nonumber\\&+&
\int_0^s \rmd t_1 \int_{t_1}^s \rmd t_2\, C^\pp(t,t_1) \sqrt{2} m(t_1) B_{TL}(t_2,t_1)
\sqrt{2} m(t_2) R^\pp(s,t_2)\,,
\eea
which can be written, using Eq.~\reff{BTL}, as
\bea
I_{18}(t,s)&=\displaystyle
\frac{6(st)^{-1/2}}{(8\pi)^{d/2}}  &\left[
\int_0^s \rmd t_2 \, t_2^2 \int_{t_2}^t \rmd \, t_1 t_1^{-1} H_L(t_1,t_2)+
s^2 \int_s^t \rmd t_2\, \int_{t_2}^t \rmd t_1 t_1^{-1}\,  H_L(t_1,t_2)
\right.\nonumber\\&&\left.+
\int_0^s \rmd t_1\, t_1^2\int_{t_1}^s \rmd t_2 t_2^{-1}\,H_L(t_2,t_1)\right]\,.
\eea
We decompose again each of the three terms in square brackets in
three parts, according to the parameterization~\reff{paraL} of $H_L$. 
Also in this case we do not report the details of the calculations, but
only the final result 
\be
I_{18}(t,s)=\frac{6 s^{1+\e/2}}{(8\pi)^{d/2}}\left(\frac{s}t\right)^{1/2}
\left[\frac2\e-\frac2\e \ln x+\frac{\ln^2 x}2+\frac32\ln x+\frac{\pi^2}3-
\frac{17}3+{\mathcal F}(x)+O(\e)
\right]\,,
\ee
where
\be
{\mathcal F}(x)=5+\frac{x}3-4\ln(1-x)-\frac2{x^2}[\ln(1-x)+x]+6\frac{\ln(1-x)}x
-2\Li_2(x)\,\label{Fx}
\ee
[${\cal F}(0) = 0$].

There is a last 1PI diagram
\bea
B_{CC}^{TL}(t,t')&=&\int (\rmd q) C^\pp_\q(t,t')C^\s_\q(t,t')\nonumber\\
&=& \int (\rmd q)
4(tt')^{-2} \rme^{-2 q^2(t+t')} \int_0^{t_<} \rmd t_1\, t_1^3 \rme^{2 q^2t_1}
\int_0^{t_<} \rmd t_2\, t_2 \rme^{2 q^2t_2}
\nonumber\\&=&
\frac{4(tt')^{-2}}{(8\pi)^{d/2}}\int_0^{t_<} \rmd t_1\int_0^{t_<} \rmd t_2
\frac{ t_1^3 t_2}{(t+t'-t_1-t_2)^{d/2}}\,,
\eea
in terms of which we express $I_{19}$ as
\bea
I_{19}&=&\int_0^s \rmd\t_1\int_0^t \rmd\t_2 \, R^\pp(s,\t_1)\sqrt2 m(\t_1)
B_{CC}^{TL}(\t_1,\t_2)\sqrt{2} m(\t_2) R^\pp(t,\t_2)\nonumber\\&=&
3(st)^{-1/2}\left[2\int_0^s \rmd\t_1\int_0^{\t_1} \rmd\t_2\, B_{CC}^{TL}(\t_1,\t_2) +
\int_0^s \rmd\t_1\int_s^t  \rmd\t_2 \, B_{CC}^{TL}(\t_1,\t_2)\right]
\nonumber\\&=&
3(st)^{-3/2}[2 B+A]\,.
\eea
$A$ and $B$ do not diverge for
$d\to4$ and therefore 
one can calculate them directly in $d=4$:
\bea
A&=&4 (8\pi)^{-d/2}\int_0^s \rmd\t_1\int_s^t \rmd\t_2 \frac1{(\t_1\t_2)^2}
\int_0^{\t_1} \rmd t_1\int_0^{\t_1} \rmd t_2
\frac{t_1^3 t_2}{(\t_1+\t_2-t_1-t_2)^2} + O(\e)
\nonumber\\&=& 4(8\pi)^{-d/2}s^2
\left[{\cal E}(s/t)-1 -\frac{\pi^2}{12}+\frac83\ln2  \right]
+ O(\e)\,,\\
B&=&4(8\pi)^{-d/2}\int_0^s \rmd\t_1\int_0^{\t_1} \rmd\t_2  \frac1{(\t_1\t_2)^2}
\int_0^{\t_2} \rmd t_1\int_0^{\t_2} \rmd t_2
\frac{t_1^3 t_2 }{(\t_1+\t_2-t_1-t_2)^2} + O(\e)
\nonumber\\&=&
4(8\pi)^{-d/2}s^2 \left(\frac{29}{48}+\frac{\pi^2}{24}-\frac43 \ln2\right)+O(\e)\; ,
\eea
with
\be
{\cal E}(x)= \frac12 \int_0^x\rmd y_1 \, (y_1^2x^{-2}-1) \int_0^1\rmd x_1
\int_0^1\rmd x_2 \frac{x_1^3 x_2}{(1 + y_1^{-1} - x_1 - x_2)^2}\,.
\label{Ex}
\ee
and ${\cal E}(0)=0$.
Thus
\be
I_{19}=\frac{12}{(8\pi)^{d/2}}s \left(\frac{s}t\right)^{1/2}
\left[\frac{5}{24}+{\cal E}(x)\right]\,.
\ee
Summing all these contributions according to Eq.~\reff{CsumP} (see
also Eq.~\reff{Cppm1loop}) one finds
that the function $f_C^\pp$ introduced in Eq.~\reff{CresultP} is given by
\be
f_C^\pp(x)=\frac{{\cal G}(x)+{\cal F}(x)+2 {\cal E}(x)}{12}\,,
\label{fcp}
\ee
where ${\cal G}(x)$, ${\cal F}(x)$, and ${\cal E}(x)$ are given by 
Eqs.~\reff{Gx}, \reff{Fx}, and \reff{Ex}, respectively.

\end{document}